\documentclass[onecolumn,showpacs,aps,prd,superscriptaddress,amsmath,amssymb,floatfix,nofootinbib]{revtex4-2}
\usepackage{graphicx}
\usepackage{amsmath}
\usepackage{bm}
\usepackage{color}
\usepackage{subcaption}

\usepackage[english]{babel}
\usepackage[utf8]{inputenc}
\usepackage{mathtools}
\usepackage{indentfirst}
\usepackage{epsfig}

\graphicspath{{Pictures/}}
\DeclareGraphicsExtensions{.pdf,.png,.jpg,.jpeg}

\usepackage{hyperref}
\hypersetup{
    colorlinks=true,
    linkcolor=blue,
    filecolor=magenta,      
    urlcolor=cyan,
}

\begin{document}

\title{Energy correlation of bottom quarks from decays of top quarks in electron--positron annihilation}
\author{Ivan V. Truten}  \email{i.truten@kipt.kharkov.ua}
\affiliation{NSC ``Kharkiv Institute of Physics and Technology'', 61108 Kharkiv, Ukraine}
\author{Alexander Yu. Korchin} \email{korchin@kipt.kharkov.ua}
\affiliation{NSC ``Kharkiv Institute of Physics and Technology'', 61108 Kharkiv, Ukraine}
\affiliation{V.N.~Karazin Kharkiv National University, 61022 Kharkiv, Ukraine}
\date{\today}

\begin{abstract}
Joint energy distribution of the bottom quark and antiquark from decays of the 
top quark and antiquark produced in the reaction $e^+ e^- \to t \bar{t}$ is studied. Main emphasis is put on $CP$-violation effects in the interaction of the photon and $Z$ boson with 
the top quarks. Energy asymmetries of $b$ and $\bar{b}$ quarks, which give access to the $CP$-violating terms, are considered. To estimate the magnitude of these asymmetries, the $CP$-violating $\gamma t t $ and 
$Z t t$ couplings are calculated in one-loop model with exchange of the Higgs boson with the mass of 125 GeV. Interaction of this boson
with the top quarks is assumed to include scalar and pseudoscalar couplings. Values of these couplings are constrained from the recent CMS analysis. 
Energy dependence of the asymmetries of $b$ and $\bar{b}$ quarks is calculated 
up to $\sqrt{s}=1.2$ TeV and some interesting features of their behavior are observed. 
These observables can be of interest for future studies at electron--positron colliders CLIC and ILC.
\end{abstract}


\maketitle

\setcounter{footnote}{0}

\section{\label{sec:Introduction}Introduction}

The main interest of this article is investigation of the $CP$ violation in the electron--positron annihilation into a pair of top quarks decaying 
into $W$ bosons and bottom quarks. Such process is planned to be explored at 
future electron--positron colliders like International Linear Collider (ILC) \cite{Behnke:2013xla, Baer:2013cma, Yamamoto:2021kig} 
and Compact Linear Collider (CLIC) \cite{Aicheler:2012bya, deBlas:2018mhx, Zarnecki:2019vrn, Zarnecki:2020ics, Kemppinen:2021urj}.
The ILC will start at the center-of-mass (CM) energy of 250 GeV followed by 500 GeV upgrade \cite{Zarnecki:2020ics, Bambade:2019fyw}. The CLIC promises to be a good candidate for production of the on-mass-shell top quark and studying its properties.
At the first construction stage of CLIC, the CM energy is planned to be 380 GeV with expected 
integrated luminosity of 1 ab$^{-1}$, which will include 100 fb$^{-1}$ collected near the $t\bar{t}$ production 
threshold \cite{Zarnecki:2019vrn, Zarnecki:2020ics, CLICdp:2018cto, Roloff:2018dqu}.
One can also mention the proposed $e^+ e^-$ Future Circular Collider FCC-ee~\cite{FCC:2018byv,FCC:2018evy}, 
which at the highest energy will be able to determine the top-quark electroweak couplings 
with a sub-percent precision. 

The process of $e^+ \, e^- \to t \bar{t}$ was studied in Refs.~\cite{Arens:1994jp, Arens:1992wh, Grzadkowski:1996kn}, 
in which the distributions of leptons $\ell^+, \, \ell^-$ produced in the decays of 
$W^+, \, W^-$ bosons were calculated. 
Authors of Refs.~\cite{ Christova:1998et, Bartl:1998nn} considered $CP$-violation effects 
by studying observables sensitive to difference of energy distribution of $b$ quark $d \sigma / dE_b $, and that of $\bar{b}$ quark $d \sigma / dE_{\bar{b}}$. 
In our paper we concentrate on the energy correlation of the bottom quark and antiquark, that is the joint two-quark
energy distribution $d^2 \sigma / dE_b \, dE_{\bar{b}}$. Such a distribution has much in common with the two-lepton energy distribution discussed in 
Ref.~\cite{Arens:1994jp}. 

The calculations in the present paper are performed using formalism developed in 
Refs.~\cite{Arens:1994jp, Kawasaki:1973hf} which allows one to find in a compact form 
distributions of secondary particles, in particular, the bottom quarks in 
$e^+ \, e^- \to t \bar{t} \to b \, \bar{b} \, W^+ W^-$.
In studying the energy distribution of $b$ and $\bar{b}$ quarks we take into account 
anomalous interactions of the photon and $Z$ boson with the top quarks.
These interactions include in addition to the $CP$-conserving anomalous magnetic dipole moment (AMDM) $\kappa$  
and anomalous weak-magnetic dipole moment (AWMDM) $\kappa_z$, the $CP$-violating electric dipole 
moment (EDM) $\tilde{\kappa}$ and weak-electric dipole moment (WEDM) $\tilde{\kappa}_z$ (see, e.g., \cite{Hollik:1998vz}). 
The terms with EDM and WEDM induce $CP$ violation~\cite{Atwood:2000tu, Zhang:2020jxw, Faroughy:2021sxk}. 
These couplings are, in general, functions of the electron--positron CM energy $\sqrt{s}$.

To estimate the magnitude of $CP$-violation effects, a one-loop model with exchange of the 
Higgs boson $h$ is applied. A similar model was considered earlier in Refs.~\cite{Bernreuther:1992dz, Chang:1993fu}, and we
update the calculation of the couplings $\kappa (s), \, \kappa_z (s), \, \tilde{\kappa} (s), \, \tilde{\kappa}_z (s)$ using the
current values of masses of the Higgs boson, $m_h = 125.18$ GeV, $Z$ boson, $m_z = 91.1876$ GeV, and other parameters of the Standard Model (SM).  
The interaction of the boson $h$ with the top quarks is assumed to be 
a mixture of the scalar ($S$) and pseudoscalar ($PS$) couplings determined 
by the parameters $\alpha$ and $\beta$, respectively.
A nonzero value of $\beta$ at nonzero $\alpha$ gives rise to a nonzero values of EDM and WEDM which 
lead to observable effects in the two-quark energy distribution.

From the two-quark energy distribution we construct energy asymmetries  
which give access to the $CP$-violating terms and study the dependence of these 
asymmetries on the $e^+ e^-$ energy. Of course, values of the parameters $\alpha$ and $\beta$ are crucial, 
and these values can be constrained based on recent analysis of the CMS collaboration \cite{CMS:2021nnc}. 

Note that a general analysis of $CP$-violation effects in the $e^+ e^- \to t \bar{t}$ process has been performed in \cite{Bernreuther:1992be}, and in the  $e^+ e^- \to \tau^+ \tau^-$ 
process in \cite{Bernreuther:1993nd}. The consideration in our paper is in line with this analysis,
and concentrates on the two-quark energy distribution and observables which can be 
of interest for studies at future $e^+e^-$ colliders.

The structure of the paper is as follows. In subsection~\ref{subsec:vertices} we discuss
the $\gamma t t$ and $Z t t$ vertices with anomalous couplings and existing constraints 
on the latter. The formalism which allows to connect the $e^+ \, e^- \to  t \bar{t}$ cross section for  
polarized top quarks to the $e^+ \, e^- \to b \, \bar{b} \, W^+ W^-$ cross section is described in subsection~\ref{subsec:cross sections}.
We present convenient expression for calculation of the joint energy distribution 
$d^2 \sigma / dE_b \, dE_{\bar{b}}$ as a function of the quark energies $E_b$ and $E_{\bar{b}}$.
In subsection~\ref{subsec:loop} a mechanism for generating EDM and WEDM,     
based on one-loop diagrams with exchange of the Higgs boson, 
is described. In section~\ref{sec:results} we present results of calculation of the
energy asymmetries of $b$ and $\bar{b}$ quarks for some values of the $h t t$ 
couplings. In section~\ref{sec:conclusions} conclusions are given. 

\section{\label{sec:formalism} Formalism for the process $e^+ \, e^- \to b \, \bar{b} \, W^+ \, W^- $}

\subsection{\label{subsec:vertices} $\gamma t t$ and $Z t t$ vertices}

\begin{figure}[tbh]
	\begin{center}
		\includegraphics[scale=0.75]{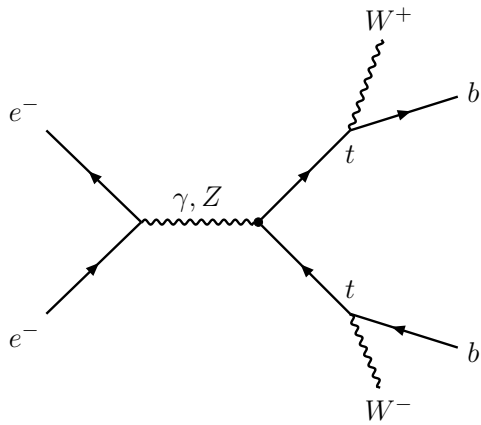}
	\end{center}
	\caption{Diagrams for $e^+ \, e^- \to t \bar{t} \to b \, \bar{b} \, W^+ W^-$ reaction. 
		The vertices $\gamma t {t}$ and $Z t {t}$ can include radiative corrections 
		and contributions beyond the SM.}
	\label{fig:1}
\end{figure}

The process is described by the tree-level diagrams in Fig.~\ref{fig:1} with the intermediate photon and 
$Z$ boson. We include the decay width in the $Z$-boson propagator. The vertices for interaction of the top quarks with intermediate particles follow from the Lagrangian for the photon 
\begin{equation}
	\mathcal{L}_{\gamma t t}=e\bar{t} \left(Q_{t}\gamma^{\mu}A_{\mu}+ \frac{1}{4m_{t}}\sigma^{\mu\nu}
	F_{\mu\nu}(\kappa+i\tilde{\kappa}\gamma_{5})\right) t, 
	\label{eq:01}
\end{equation}
and for the $Z$ boson 
\begin{eqnarray} \mathcal{L}_{Z t t} &=& \frac{g}{2\cos{\theta_w}}\bar{t} \biggl( \gamma^{\mu}
	Z_{\mu} (v_t -a_t \gamma_5) + \frac{1}{4m_{t}}\sigma^{\mu\nu}Z_{\mu\nu}(\kappa_z+i \tilde{\kappa}_z\gamma_{5})\biggr)  t .
	\label{eq:02}
\end{eqnarray}
Here $e$ is the positron charge, $g = e / \sin \theta_w$ with $\theta_w$ denoting the weak mixing angle, 
$Q_t =2/3$, the top-quark vector and axial-vector couplings are $v_t= 1/2 -4/3 \sin^2 \theta_w$ and 
$a_t=1/2$, respectively, and $m_t$ is the mass of the top quark.

The terms proportional to couplings $\kappa$ and $\kappa_z$ determine $CP$-even interaction, while 
the terms with $\tilde{\kappa}$ and $\tilde{\kappa_z}$ determine $CP$-odd interaction. 
In our previous papers~\cite{Truten:2021nkj, Truten:2019sqb} 
we studied only effects due to anomalous couplings $\kappa$ and $\kappa_z$, 
thus neglecting $CP$ violation. Here we include all the couplings in (\ref{eq:01}) and (\ref{eq:02}).
In addition, once the final-state interaction, or re-scattering effects, are taken into account, these couplings 
acquire imaginary part. In this paper we consider complex values of the couplings, and their calculation is presented in subsection~\ref{subsec:loop}.

The corresponding $\gamma t {t}$ and $Z t {t}$ vertices are written as
\begin{eqnarray}
	\Gamma_{\gamma t {t}}^\mu &=& -i e  
	\Big[ Q_t \gamma^\mu + i \frac{ \sigma^{\mu \nu} q_\nu}{2 m_t} (\kappa + i \tilde{\kappa}  \gamma_5) \Big], 
	\label{eq:020_vertex_g} \\ 
	\Gamma_{Z t {t}}^\mu &=& -i \frac{g}{2 \cos \theta_w} \Big[ \gamma^\mu (v_t-a_t \gamma_5) + i  \frac{\sigma^{\mu \nu} q_\nu}{2 m_t} 
	(\kappa_z + i \tilde{\kappa}_z \gamma_5) \Big], 
	\label{eq:020_vertex_Z}
\end{eqnarray}
where $q^\nu= k^\nu + k^{\prime \nu}$ is the four-momentum of the intermediate photon ($Z$ boson). 
The couplings $\kappa, \, \kappa_z, \, \tilde{\kappa}, \, \tilde{\kappa_z}$ are functions of $s=q^2$.

At present there are no unique constraints on values
of the anomalous couplings in Eqs.~(\ref{eq:020_vertex_g}) and (\ref{eq:020_vertex_Z}). 
An overview of the global analysis of the couplings is given in \cite{Bouzas:2021gwx}. 
The results are presented in the form of bounds on the Wilson coefficients $C_{uB }^{33}$ and $C_{uW}^{33}$ associated with 
the dimension-6 operators ${\cal O}_{uB }^{33}$ and ${\cal O}_{uW}^{33}$ respectively, in the SM 
effective Lagrangian \cite{Grzadkowski:2010es}. 
These Wilson coefficients\footnote{Sometimes in the literature another notation for these Wilson coefficients is used: 
$ C_{tW} = {C}_{uW}^{33} $ and $C_{tB} = {C}_{uB }^{33} $.} 
are related to the anomalous couplings through
\begin{eqnarray}
&& \kappa = \sqrt{2 } \, \frac{m_t}{m_z} \,
\left( \sin \theta_w \, \mathrm{Re} \, \bar{C}_{uW}^{33} +\cos \theta_w \, \mathrm{Re} \, \bar{C}_{uB }^{33}\right) 
\frac{1}{ \cos \theta_w \sin \theta_w }, \label{eq:03} \\
 && \kappa_z = 2 \sqrt{2} \, \frac{m_t}{m_z} \, \left( \cos \theta_w \, \mathrm{Re} \, \bar{C}_{uW}^{33} - 
\sin \theta_w \, \mathrm{Re} \, \bar{C}_{uB }^{33} \right), 
\label{eq:04}
\end{eqnarray}
where $\bar{C}_{uW}^{33} \equiv \left( v^2 / \Lambda^2 \right) \, {C}_{uW}^{33} $, \
 $\bar{C}_{uB }^{33} \equiv \left( v^2 / \Lambda^2 \right) \, {C}_{uB }^{33} $, \
$v=\left(\sqrt{2}G_{\rm F}\right)^{-1/2}\approx 246$ GeV is vacuum expectation value of 
the Higgs field, $G_F = 1.1663787(6) \times 10^{-5}$ GeV$^{-2}$~\cite{Zyla:2020zbs},
and $\Lambda$ is the ``new physics'' scale which is usually chosen 1 TeV.     
The couplings $\tilde{\kappa}$ and $\tilde{\kappa}_z$ are related to the imaginary part 
of these Wilson coefficients by the relations similar to (\ref{eq:03}) and (\ref{eq:04}).

Several groups \cite{Ethier:2021bye, Brown:2020sjx, Ellis:2020unq, Miralles:2021dyw, Bissmann:2020mfi} 
recently reported bounds on the real part of $C_{uB }^{33}$ and $C_{uW}^{33}$ which follow from 
the experimental branching ratio of the $\bar{B} \to X_s \gamma$ decays, and from the LHC cross sections of the single $t$-quark production, 
$t \bar{t}$, $t \bar{t} \gamma$, and also $W$-boson polarization in the $t$-quark decay. The strongest bounds are obtained by the EFTfitter group~\cite{Bissmann:2020mfi} from a fit of coefficients to the top-quark data 
at the LHC and combined dataset of the $B$-physics and $Zbb$ data. Using these results we get
\begin{equation}
\kappa = ( -0.24 , \, 0.28 ), \qquad  \qquad \kappa_z = (-0.15, \, 0.32). 
\label{eq:05}
\end{equation}

The wider limits on the couplings are obtained by the Fitmaker group~\cite{Ellis:2020unq}. 
Using these constraints (for ``individual 95\% C.L. range'') we have    
\begin{equation}
\kappa = ( -1.54 , \, 0.5 ), \qquad  \qquad \kappa_z = (-0.22, \, 0.85). 
\label{eq:06}
\end{equation}
In calculations in section \ref{sec:results} we use values of $\kappa$ and $\kappa_z$ which do not contradict the bounds (\ref{eq:05}). 
At the same time there are no constraints on the imaginary part of the Wilson coefficients and therefore for the couplings 
$\tilde{\kappa}$ and $\tilde{\kappa}_z$ we rely on a model in subsection~\ref{subsec:loop}.

\subsection{\label{subsec:cross sections} Cross sections}

\begin{figure}[tbh]
	\begin{center}
		\includegraphics[scale=0.9]{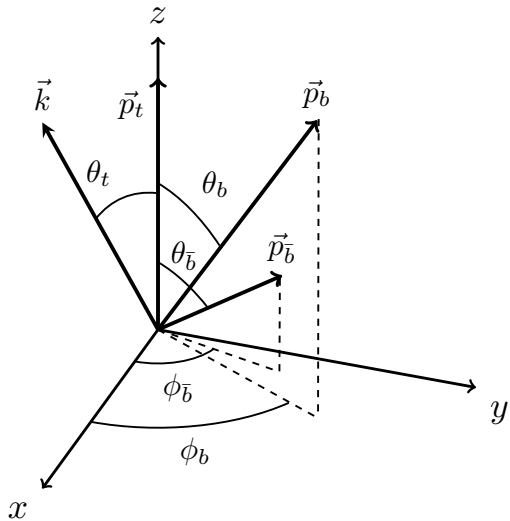}
	\end{center}
	\caption{The coordinate system used in definition of the particle momenta.}
	\label{fig:2}
\end{figure}

In calculation of the cross sections in the CM frame it is convenient to use the coordinate 
system in which the top-quark and antiquark momenta are directed along the $OZ$ axis, 
the electron and positron momenta lie in the plane $XOZ$, the bottom-quark and antiquark momenta 
point in arbitrary directions (see Fig.~\ref{fig:2}). Further, $\theta_t$ is the angle between the electron and the top-quark momenta. The polar and azimuthal angles of $b$ and $\bar{b}$ quarks are defined in Fig.~\ref{fig:2}.
In this system the four-momenta of $e^-$, $e^+$, $t$, $\bar{t}$, $b$ and $\bar{b}$ are, respectively
\begin{eqnarray}
	k^\mu &=& E \bigl(1, \,  \sin{\theta_t}, \, 0, \, \cos{\theta_t} \bigr), \nonumber \\
	k^\prime{}^\mu &=& E \bigl(1, \, -\sin{\theta_t}, \, 0, \, -\cos{\theta_t} \bigr), \nonumber \\
	p_t^\mu &=& \bigl(E_t,\, 0,\, 0,\, p_t \bigr), \nonumber \\
	{p^\prime_t}^\mu &=& \bigl( E_t,\, 0,\, 0,\, - p_t \bigr ), \nonumber \\
	p_b^\mu &=& E_b \bigl( 1, \, \sin{\theta_b} \cos{\phi_b}, \, \sin{\theta_b} \sin{\phi_b}, \, \cos{\theta_b} \bigr), \nonumber \\
	p_{\bar{b}}^\mu &=& E_{\bar{b}} \bigl( 1, \, \sin \theta_{\bar{b}} \cos{\phi_{\bar{b}}}, \, \sin \theta_{\bar{b}} \sin{\phi_{\bar{b}}}, 
	\, \cos \theta_{\bar{b}} \bigr), 
	\label{eq:011}
\end{eqnarray}
where $E_t = E = \sqrt{s}/2$ is the energy of the top quark (antiquark), electron (positron), 
$p_t=\sqrt{E_t^2-m_t^2}$. The mass of the bottom quark (antiquark) is neglected.

The evaluation of cross sections of the two-step processes is based on the formalism of 
Refs.~\cite{Kawasaki:1973hf, Arens:1994jp} in which 
bottom quark and antiquark are produced from the decays of the on-mass-shell top quark and antiquark. 
Note that the single tree-level diagram with the photon exchange satisfies the gauge invariance, as shown 
explicitly in \cite{Truten:2021nkj}. Using results of \cite{Truten:2021nkj} the joint energy 
distribution of $b$ and $\bar{b}$ quarks in the $e^+e^- \to b \, \bar{b}\, W^+ \, W^-$ reaction 
can be written as
\begin{equation}
	\frac{d^2 \sigma_{e^+e^- \to b \, \bar{b} \, W^+ \, W^-}}{d E_b \, d E_{\bar{b}}} = 4 \left( \frac{m_t}{4 \pi p_t p_b^0 } \right)^2 
	\int \frac{d \sigma_{e^+ e^- \to t \, \bar{t}} \, (n^\mu , \, {n^\prime}^\mu)}{d \Omega_t} \, 
	d \Omega_t \, d \phi_b \, d \phi_{\bar{b}}.
	\label{eq:1011}
\end{equation}

In this equation the $b$-quark polar angle is fixed to 
\begin{equation}
	\cos \theta_b = \frac{m_w^2 - m_t^2 + 2 E_t E_b}{2 p_t E_b},
	\label{eq:1005}
\end{equation}
while the $\bar{b}$-quark polar angle is given by 
\begin{equation}
	\cos \theta_{\bar{b}} = - \frac{m_w^2 - m_t^2 + 2 E_t E_{\bar{b}}}{2 p_t E_{\bar{b}}}. 
	\label{eq:1007}
\end{equation}
In Eq.~(\ref{eq:1011}) $p_b^0 = (m_t^2 -m_w^2)/(2 m_t)$ is the energy of $b$ quark (antiquark) 
in the rest frame of the decaying $t$ quark (antiquark), and $m_w$ is the mass of $W$ boson. 

The four-vectors $n^\mu$ and ${n^{\prime}}^\mu$ in (\ref{eq:1011}) denote the top quark and antiquark polarizations. 
They are defined as
\begin{equation}
	n^\mu = \alpha_b \left(- \frac{p_t^\mu}{m_t} + \frac{m_t p_b^\mu}{p_t \cdot p_b} \right), \qquad \quad
	{n^\prime}^\mu = \alpha_{\bar{b}} \left(- \frac{p^{\prime \mu}_t}{m_t} + \frac{m_t p_{\bar{b}}^\mu}{p^\prime_t \cdot p_{\bar{b}}} \right).
	\label{eq:1012}
\end{equation}
Here the parameters $\alpha_b = (2m_w^2 - m_t^2)/(2m_w^2 + m_t^2)$ and $\alpha_{\bar{b}} = - \alpha_b$ determine the asymmetry 
in the decays of the top quark and antiquark, respectively. The four-vectors in (\ref{eq:1012}) satisfy the conditions:
$n \cdot p_t = n^{\prime} \cdot p^{\prime}_t =0$, \ $n \cdot n = - \alpha_b^2 $ and $n^\prime \cdot n^\prime = - \alpha_{\bar{b}}^2 $.

The energies of $b$ and $\bar{b}$ quarks are confined within the limits
\begin{equation}
	E_{-} \le (E_b, \, E_{\bar{b}})  \le E_{+}, \qquad \qquad 
	E_{\pm}=\frac{m_t \, p_b^0}{E_t \mp p_t }= \frac{\sqrt{s}(1 \pm V) p_b^0 }{2 m_t},
	\label{eq:1006}
\end{equation}
where $V = \sqrt{1 - 4 m_t^2/s}$ is the velocity of the top quark.

From the two-particle distribution (\ref{eq:1011}) one can derive the energy distribution 
of a single $b$ quark in the process $e^+e^- \to b \, W^+ \, \bar{t} $. 
By integrating (\ref{eq:1011}) over the energy of the $\bar{b}$ quark we obtain
\begin{equation}
	\frac{d \sigma_{e^+e^- \to b \, W^+  \bar{t}}}{d E_b} = 
	2 \left( \frac{m_t}{4 \pi p_t  p_b^0 } \right) \int \frac{d \sigma_{e^+ e^- \to t \, \bar{t}} \, (n^\mu , \, 0)}{d \Omega_t} \,  d \Omega_t \, d \phi_b.
	\label{eq:1009}
\end{equation}

The analogous procedure leads to the energy distribution of $\bar{b}$ quark  in $e^+e^- \to \bar{b} \, W^- \, t $:
\begin{equation}
	\frac{d \sigma_{e^+e^- \to   \bar{b} \, W^- t}}{d E_{\bar{b}}}  = 2 \left( \frac{m_t}{4 \pi p_t p_b^0 } \right) \int \frac{d \sigma_{e^+ e^- \to 
			t \, \bar{t}} \, (0, \, {n^\prime}^\mu) }{d \Omega_t} \, d \Omega_t \, d \phi_{\bar{b}}
	\label{eq:1010}
\end{equation}

Eqs.~(\ref{eq:1011}), (\ref{eq:1009}) and (\ref{eq:1010}) allow one to find the cross sections for arbitrary model of the underlying 
process $e^+ e^- \to t \bar{t}$. 


\subsection{\label{subsec:loop} Couplings in one-loop model}

\begin{figure}[tbh]
	\centering
	\begin{subfigure}[b]{0.32\textwidth}
		\includegraphics[width=\textwidth]{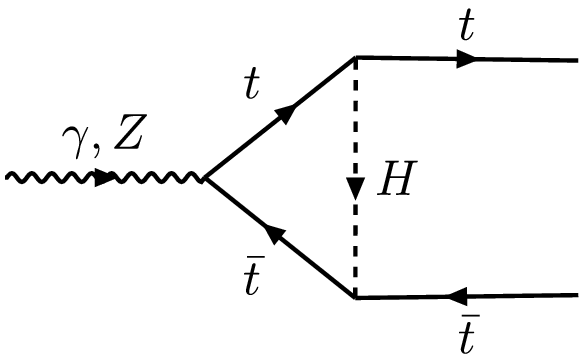}
		\caption{}
	\end{subfigure}
	\begin{subfigure}[b]{0.32\textwidth}
		\includegraphics[width=\textwidth]{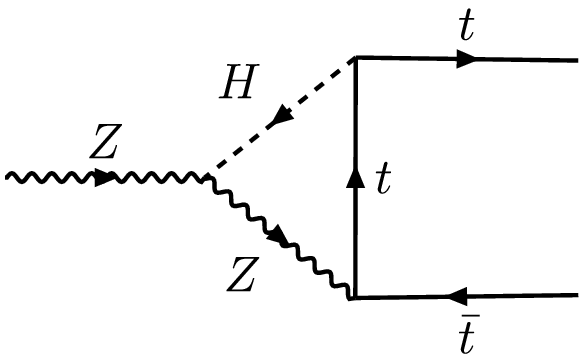}
		\caption{}
	\end{subfigure}
	\begin{subfigure}[b]{0.32\textwidth}
		\includegraphics[width=\textwidth]{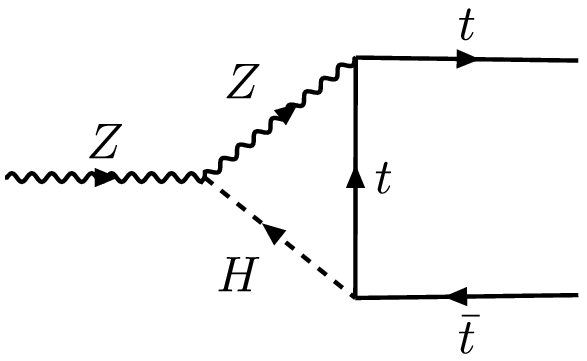}
		\caption{}
	\end{subfigure}
	\caption{The $\gamma t t $ and $Z t t$ vertex corrections. Wavy lines denote photon or $Z$ boson, 
		dashed line -- Higgs boson, solid lines -- top quark.}
	\label{fig:5}
\end{figure}

For an estimation of the magnitude of the couplings $\kappa$, $\kappa_z$, $\tilde{\kappa}$ and $\tilde{\kappa}_z$  
we use the $\gamma t {t} $ and $Z t {t}$ vertex corrections shown in Fig.~\ref{fig:5}, which include exchange of the intermediate Higgs boson. 
Note that a similar model has been used in Ref.~\cite{Chang:1993fu} which considered 
energy distribution of leptons $\ell^+$ and $\ell^-$ coming from $W^+$ and 
$W^-$ decays in $e^+ e^- \to t \bar{t}$ process. In \cite{Bernreuther:1992dz} a
more general two-Higgs doublet model has been considered.

In order to generate $CP$-violation effects in Eqs.~(\ref{eq:020_vertex_g}) and 
(\ref{eq:020_vertex_Z}), the interaction of the Higgs boson $h$ with the top quarks is chosen in the form 
\begin{equation}
	{\cal L}_{h t t}=- \frac{m_t}{v}\,h\,{\bar t } \left(\alpha + i \, \beta \gamma_5\right) t \,,
	\label{eq:1013}
\end{equation}
which includes scalar ($S$) and pseudoscalar ($PS$) parts, where $\alpha$ and $\beta$ are real-valued parameters. 
Values $\alpha=1$ and $\beta=0$ correspond to the SM. 
A nonzero coupling $\beta$ gives rise to structures proportional to $\gamma_5 $ 
in (\ref{eq:020_vertex_g}) and (\ref{eq:020_vertex_Z}). 
One can consider Eq.~(\ref{eq:1013}) as a phenomenological parameterization 
of effects of physics beyond the SM (BSM). Such an interaction has been used in various 
papers (see, for example, \cite{Korchin:2013ifa, Korchin:2014kha, Korchin:2016rsf, Chen:2017nxp}).

The Higgs-boson interaction with $Z$ bosons is taken the same as in the SM, i.e. $ (m_z^2 /v) \, h \, Z^\mu Z_\mu $. In general, the $hZZ$ vertex BSM can include additional structures, in particular, a term 
$\sim \epsilon^{\mu \nu \rho \sigma} h \, \partial_\mu Z_\nu \partial_\rho Z_\sigma $ which accounts for 
the negative $CP$ parity (here $\epsilon^{\mu \nu \rho \sigma}$ is the antisymmetric Levi-Civita symbol). 
In the present paper we neglect this and other anomalous $h ZZ$ terms in calculation of the diagrams `b' and `c' in Fig.~\ref{fig:5}.

Note that the couplings in (\ref{eq:020_vertex_g}) and (\ref{eq:020_vertex_Z}) 
acquire imaginary parts due to absorptive contribution in the diagrams in Fig.~\ref{fig:5}. 
Using the Cutkosky rules one obtains for the $\gamma t {t}$ vertex in  Fig.~\ref{fig:5} (diagram `a')
\begin{equation}
	{\rm Im}\, \tilde{\kappa} (s) = \alpha \beta \frac{Q_t m_t^4 }{8 \pi E_t p_t v^2} 
	\left( 1 - \frac{m_h^2}{4 p_t^2} \log \frac{m_h^2+4 p_t^2}{m_h^2} \right) \, \theta(s - 4 m_t^2),
	\label{eq:1014}
\end{equation}
where $\theta(x)$ is the Heaviside step function.

For the $Z t {t}$ vertex in Fig.~\ref{fig:5} (diagram `a') we find
\begin{equation}
	{\rm Im}\, \tilde{\kappa}_z (s)_{a} = \frac{v_t}{Q_t} {\rm Im}\, \tilde{\kappa} (s).
	\label{eq:1015}
\end{equation}
The contribution to the $Z t {t}$ vertex from diagrams `b' and `c' in Fig.~\ref{fig:5} is more complicated. 
It is equal to 
\begin{eqnarray}
	{\rm Im}\, \tilde{\kappa}_z  (s)_{b, \, c} &=& \beta \frac{v_t m_t^2 m_z^2 }{8 \pi E_t p_t v^2} 
	\left[ - \frac{k_z}{p_t} +L \left(1- \frac{E_z}{2E_t} + \frac{m_z^2- 2 E_t E_z}{4 p_t^2} \right) \right] \nonumber \\
	&& \times \theta(s - (m_h +m_z)^2) 
	\label{eq:1016}
\end{eqnarray}
with definitions
\begin{eqnarray}
	E_z & =& E_t - \frac{m_h^2- m_z^2}{4 E_t}, 
	\qquad k_z = \sqrt{E_z^2 - m_z^2},  \nonumber \\ 
	L &=& \log \left( \frac{ 2 E_t E_z - 2 p_t k_z -m_z^2 }{2E_t E_z + 2 p_t k_z -m_z^2 } \right).
	\label{eq:1017}
\end{eqnarray} 

The real part of $\tilde{\kappa} (s)$ and $\tilde{\kappa}_z (s)$ can be calculated using the dispersion 
relations (see {\rm e.g.} \cite{Chang:1993fu})
\begin{eqnarray}
	&& {\rm Re} \, \tilde{\kappa} (s) =  \frac{1}{\pi} {\rm PV} \int_{s_0}^\infty \frac{{\rm Im}\, \tilde{\kappa} (s') }{s' -s} d s'  , \nonumber \\
	&& {\rm Re} \, \tilde{\kappa}_z (s) = \frac{1}{\pi} {\rm PV}  \int_{s_1}^\infty \frac{{\rm Im}\, \tilde{\kappa}_z (s') }{s' -s} d s',
	\label{eq:1018}
\end{eqnarray}
where symbol `PV' means principle value, $s_0 = 4 m_t^2$, and $s_1 = 4 m_t^2$ for the intermediate top quarks 
(Fig.~\ref{fig:5} `a') and $s_1= (m_h+m_z)^2$ for the intermediate $Z$ and $h$ bosons (Fig.~\ref{fig:5} `b' and `c' ). 
Otherwise, the real parts can be calculated directly from the diagrams in Fig.~\ref{fig:5}.
The couplings $\kappa (s)$ and $\kappa_z (s)$ are obtained in a similar way.
Note that our results for $ \tilde{\kappa} (s)$ and $ \tilde{\kappa}_z (s)$ are in agreement 
with results of Refs.~\cite{Bernreuther:1992dz, Chang:1993fu}.

\begin{figure}[tbh]
	\begin{center}
		\includegraphics[scale=1]{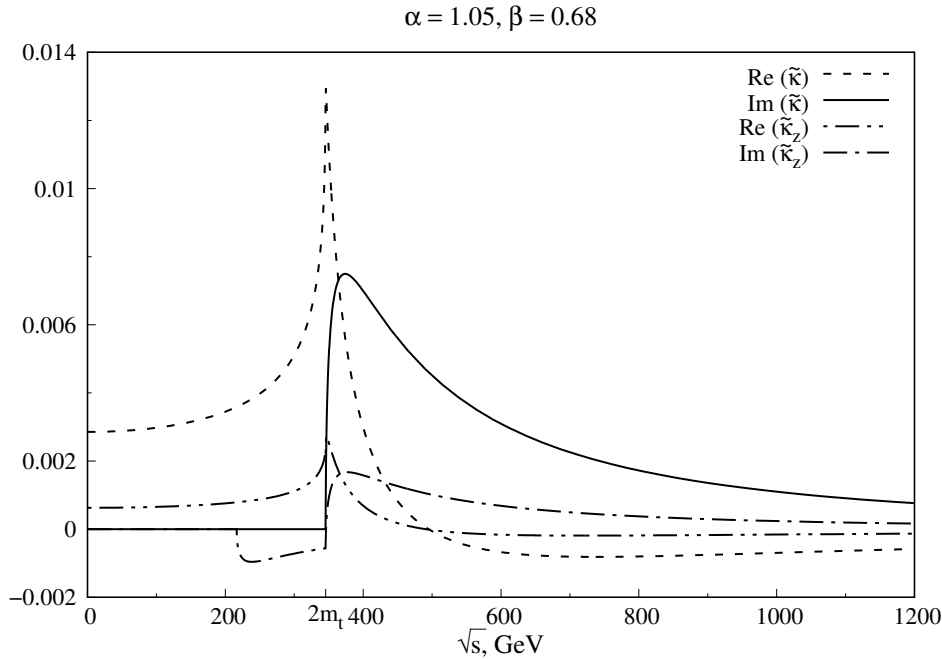}
	\end{center}
	\caption{Energy dependence of the real and imaginary parts of $\tilde{\kappa} (s)$ 
		and $\tilde{\kappa}_z (s)$.}
	\label{fig:couplings}
\end{figure}

We calculated numerically the couplings for the values of masses of the top quark, Higgs boson and $Z$ boson 
from PDG \cite{Zyla:2020zbs}. In Fig.~\ref{fig:couplings} the energy dependence of the real and imaginary parts of 
$\tilde{\kappa} (s)$ and $\tilde{\kappa}_z (s)$ is shown for the values $\alpha=1.05$ and $\beta=0.68$ (the choice
of these values is explained after Eqs.~(\ref{eq:1029a}) and (\ref{eq:1029b})).

For example, at the energy $\sqrt{s} = 380$ GeV, relevant for the collider CLIC, 
the couplings $\kappa, \, \tilde{\kappa}, \, \kappa_z, \, \tilde{\kappa}_z$ for arbitrary values of $\alpha$ and $\beta$ are 
\begin{eqnarray} 
	\kappa &=& \alpha^2 \, (0.0047 + i \, 0.0060) - \beta^2 \, (0.0020 + i \, 0.0044), \nonumber \\
	\tilde{\kappa} &=& \alpha \beta\, (0.0068 + i \, 0.0104 ),  \nonumber \\
	\kappa_z &=& \alpha^2 \,(0.0014 + i \, 0.0017) - \alpha \, (0.0004 + i \, 0.0007) - \beta^2 \, (0.0006 + i \,0.0013), \nonumber \\
	\tilde{\kappa}_z &=& \alpha \beta \, (0.0019 + i \, 0.0030 ) - \beta \, (0.0004 + i \, 0.0007) .
	\label{eq:1019}
\end{eqnarray}

\section{\label{sec:results} Energy correlation and asymmetries of bottom quark and antiquark}

As pointed out in Ref.~\cite{Chang:1993fu}, the $CP$ violation shows up in the energy asymmetries of secondary 
leptons, which arise if the top-quark configurations $t_L \bar{t}_L$ and $t_R \bar{t}_R$ in $e^+ e^- \to t \bar{t}$ 
appear with different probabilities. In this section we study the energy asymmetries of the bottom quark and antiquark 
which appear due to the couplings $\tilde{\kappa} (s)$ and $\tilde{\kappa}_z (s)$. 

Indeed, analysis of transformation properties of Lagrangians (\ref{eq:01}) and (\ref{eq:02}) shows that 
the energy asymmetries of $b$ and $\bar{b}$ quarks arises due to the terms which are $CP$-odd and 
$T$-even, and thus $CPT$-odd. Therefore the following combinations of the couplings lead 
to the asymmetries: 
$ \, Q \, {\rm Im} \, \tilde{\kappa},  \, v_t \, {\rm Im} \, \tilde{\kappa}_z,  \, 
{\rm Im} \, \tilde{\kappa} \, {\rm Re} \, \kappa, \, {\rm Re} \,  \tilde{\kappa} \, {\rm Im} \, \kappa, 
\, {\rm Im} \, \tilde{\kappa}_z \, {\rm Re} \, \kappa_z, \, {\rm Re} \, 
\tilde{\kappa}_z \, {\rm Im} \, \kappa_z, \ldots $ 
Besides, the amplitude in Fig.~\ref{fig:1} is complex due to the $Z$-boson propagator, and 
the terms proportional to $\, \Gamma_z  \, {\rm Re} \, \tilde{\kappa}$ 
and $\Gamma_z \, {\rm Re} \, \tilde{\kappa}_z$
also contribute to the energy asymmetries (here $\Gamma_z$ is the $Z$-boson decay width). 
As is seen from Fig.~\ref{fig:couplings}, the values of all couplings are small and thus 
it is sufficient to keep only linear in the couplings terms proportional to 
${\rm Im} \, \tilde{\kappa}$, ${\rm Im} \, \tilde{\kappa}_z$, 
${\rm Re} \, \tilde{\kappa}$ and ${\rm Re} \, \tilde{\kappa}_z$.

To extract information on the $CP$-violating terms we analyze the structure of the two-quark cross section 
(\ref{eq:1011}). Let us define the distribution
\begin{equation}
W(E_b, \, E_{\bar{b}}) = \frac{1}{\sigma_0} \, \frac{d^2 \sigma_{e^+e^- \to b \, \bar{b} \, 
		W^+ \, W^-}}{d E_b \, d E_{\bar{b}}} 
\label{eq:1020}
\end{equation}
where $\sigma_0 \equiv \sigma_{e^+e^- \to t \, \bar{t}} $ is the total cross section 
of the $e^+e^- \to t \, \bar{t}$ process. 
For brevity we denote further $\varepsilon \equiv E_b$ and $ \bar{\varepsilon} \equiv E_{\bar{b}}$. 

The distribution $W(\varepsilon, \, \bar{\varepsilon})$ is normalized to unity
\begin{equation}
\int_{E_-}^{E_+} \int_{E_-}^{E_+}  W(\varepsilon, \, \bar{\varepsilon}) \,  d \varepsilon \, d \bar{\varepsilon}  =1. 
\label{eq:1021}
\end{equation}

\begin{figure}[tbh]
	\begin{center}
		\includegraphics[scale=1]{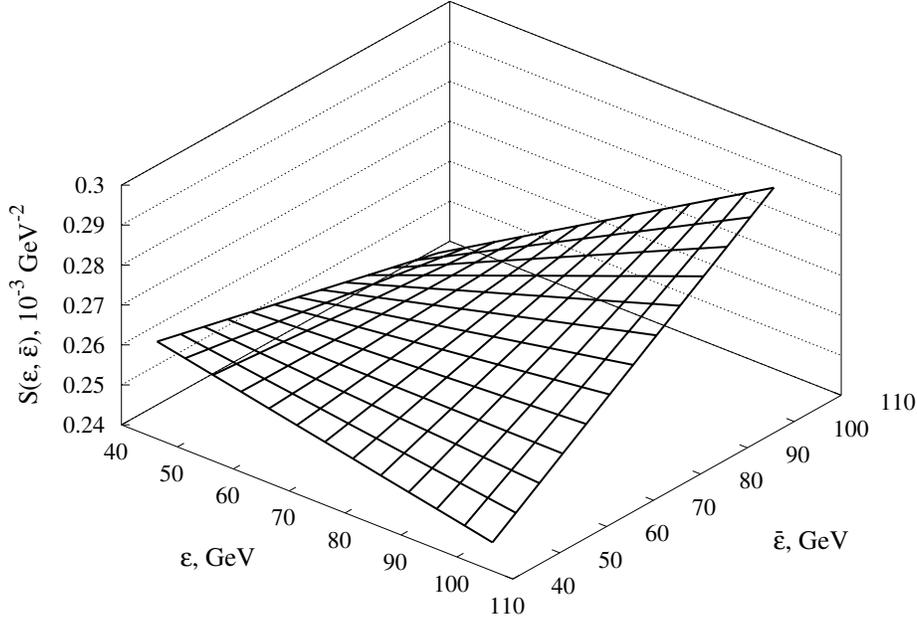}
	\end{center}
	\caption{Symmetrical function $S (\varepsilon, \, \bar{\varepsilon})$ at $\sqrt{s}=380$ GeV. }
	\label{fig:S_sym}
\end{figure}

The general structure of this distribution is as follows 
\begin{eqnarray}
W(\varepsilon, \, \bar{\varepsilon}) &=& S (\varepsilon, \, \bar{\varepsilon}) + 
(\varepsilon -\bar{\varepsilon}) \, T(s), 
\label{eq:1022} \\
T(s) &=&\left[ a(s) \, {\rm Im} \, \tilde{\kappa}(s) + b(s) \, {\rm Im} \, \tilde{\kappa}_z(s) 
+ c(s) {\rm Re} \, \tilde{\kappa}(s) + d(s) {\rm Re} \, \tilde{\kappa}_z(s) \right],
\label{eq:1022a}
\end{eqnarray}
where $S (\varepsilon, \, \bar{\varepsilon})$ is the symmetrical function of the $b$ quark and antiquark 
energies $\varepsilon$ and $\bar{\varepsilon}$. As an example, it is shown in 
Fig.~\ref{fig:S_sym} for the $e^+ e^-$ energy 380 GeV.

The coefficients $a(s), \, b(s), \, c(s), \, d(s) $ in Eq.~(\ref{eq:1022a}) are functions of the $e^+e^-$ energy. 
Analysis shows that, in general, $|c(s)| \ll |a(s)|$ and $|d(s)| \ll |b(s)|$. 
For example, the dominant coefficients $a(s)$ and $b(s)$ have the form 
\begin{eqnarray}
a(s) &=& \frac{1}{\sigma_0} \, \frac{\pi \alpha^2 m_t^6 \alpha_b \left(v_e v_t \left(m_z^2 - s\right) s + 
	4 Q_t \cos^2 \theta_w \sin^2 \theta_w \left(\left(s - m_z^2\right)^2 + m_z^2 \Gamma_z^2\right)\right)}
{s^{\frac{5}{2}} V \cos^2 \theta_w \sin^2 \theta_w \left(m_t^2 - m_w^2\right)^3 \left(\left(s - m_z^2\right)^2 + m_z^2 \Gamma_z^2\right)},
\label{eq:1023}
\\ \nonumber \\
b(s) &=& \frac{1}{\sigma_0} \, \frac{4 \pi \alpha^2 m_t^6 \alpha_b \left(v_t \left(a_e^2 + v_e^2\right) s + 
	4 Q_t v_e \cos^2 \theta_w \sin^2 \theta_w \left(m_z^2 - s\right)\right)}
{s^{\frac{3}{2}} V \cos^4 \theta_w \sin^4 \theta_w \left(m_t^2 - m_w^2\right)^3 \left(\left(s - m_z^2\right)^2 + m_z^2 \Gamma_z^2\right)}. 
\label{eq:1024}
\end{eqnarray} 
Note that these coefficients depend on $\kappa(s)$ and $\kappa_z(s)$ through  $\sigma_0$.

The structure in Eq.~(\ref{eq:1022}) allows one to construct observables suitable to determine 
$T(s)$ in (\ref{eq:1022a}). The dimensionless quantity
\begin{equation} 
A (s) \equiv \int_{E_-}^{E_+} \int_{E_-}^{E_+} W(\varepsilon, \, \bar{\varepsilon}) \, 
[\theta(\varepsilon - \bar{\varepsilon}) - 
\theta(\bar{\varepsilon} - \varepsilon)] \, d \varepsilon \, d \bar{\varepsilon} 
\label{eq:1025}
\end{equation}
has the meaning of asymmetry in the energy distribution of $b$ and $\bar{b}$ quarks. 
It is equal to experimentally measured quantity $ (N_{ E_b > E_{\bar{b}}} - N _{ E_{\bar{b}} > E_b} ) /N_{tot}$ 
in terms of the number of events. 

For any function $S (\varepsilon, \, \bar{\varepsilon})$ we find 
\begin{equation}
A (s) = \frac{1}{3} (E_+ - E_-)^3 \,  T(s)
\label{eq:1026}
\end{equation}
with $E_+ - E_- = \sqrt{s} V p_b^0 /m_t$. 

Another observable, which gives information on $\tilde{\kappa}$ and $\tilde{\kappa}_z$, 
is the difference of mean energies of $b$ and $\bar{b}$ quarks
\begin{equation}
\left\langle \varepsilon \right\rangle   - \left\langle \bar{\varepsilon} \right\rangle \equiv 
\int_{E_-}^{E_+} \int_{E_-}^{E_+} 
W(\varepsilon, \, \bar{\varepsilon}) \,  (\varepsilon - \bar{\varepsilon}) \, d \varepsilon \, 
d \bar{\varepsilon} = \frac{1}{6}(E_{+}  - E_{-})^4 
\, T(s). 
\label{eq:1027} 
\end{equation}

The observables (\ref{eq:1025}) and (\ref{eq:1027}) give access to the $CP$-violating coefficients. 
More complicated asymmetries can also be constructed, for example, the difference of the $n$-th moments ($n \ge 1$)
of the distribution $W(\varepsilon, \, \bar{\varepsilon}) $, that at high energies behaves as 
\begin{equation}
\left\langle \varepsilon^n  \right\rangle - \left\langle \bar{\varepsilon}^n \right\rangle 
\sim s^{(n+3)/2} \, T(s). 
\label{eq:1028a}
\end{equation} 

\begin{figure}[tbh]
	\begin{center}
		\includegraphics[scale=0.61]{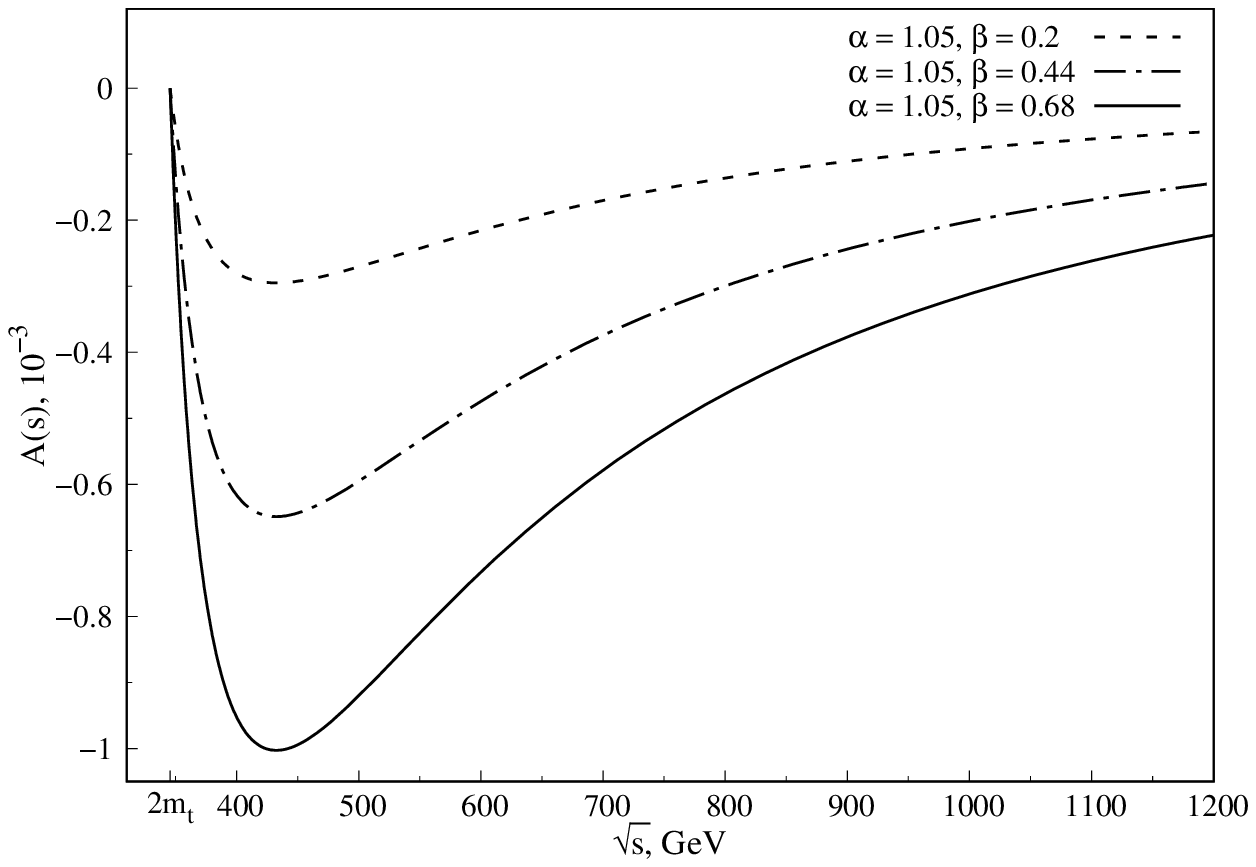}
		\includegraphics[scale=0.61]{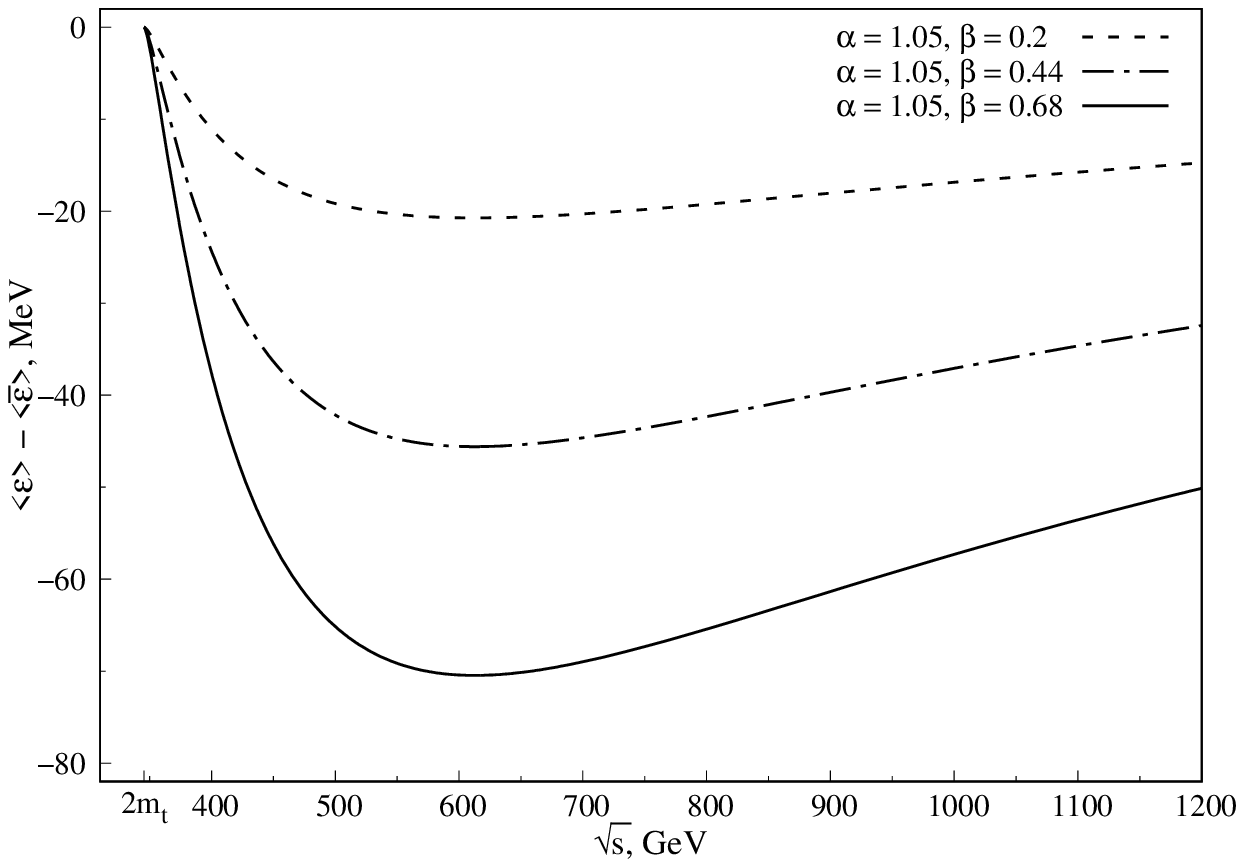}
	\end{center}
	\caption{The bottom quark--antiquark asymmetry $A(s)$ (left pannel) and  
mean-energy asymmetry $ \left\langle \varepsilon  \right\rangle   - \left\langle \bar{\varepsilon} \right\rangle$ 
(right pannel) as functions of the invariant $e^+ e^-$ energy  $\sqrt{s} = 2E$.}
	\label{fig:A(s)_Eb-Ebbar}
\end{figure}

The asymmetries (\ref{eq:1025}) and (\ref{eq:1027}) are shown in Fig.~\ref{fig:A(s)_Eb-Ebbar} for a few values 
of the parameters $\alpha$ and $\beta$ in the $h t t$ interaction (\ref{eq:1013}). The CMS Collaboration~\cite{CMS:2021nnc} 
reported the following constraints on these couplings:
\begin{eqnarray}
\alpha &=& 1.05^{+0.25}_{-0.20}, \qquad \qquad \beta = -0.01^{+0.69}_{-0.67}  \qquad {\rm (observed)}, 
\label{eq:1029a} \\
\alpha &=& 1.00^{+0.34}_{-0.26}, \qquad \qquad \beta = -0.00^{+0.71}_{-0.71}  \qquad {\rm (expected)}. 
\label{eq:1029b}
\end{eqnarray} 
The term `expected' means values obtained from the Monte Carlo simulation, while 
`observed' stands for values obtained by the CMS \cite{CMS:2021nnc} from measurements. 
The CMS extracted the $htt$ couplings, as well as other Higgs-boson couplings ($hgg, \, hVV$), based on a combined analysis of the Higgs 
production cross sections via the mechanisms of the gluon fusion $gg \to h$, production in association with a top-quark pair $gg \to t \bar{t} h$ and with 
a single top quark $b q \to t q^\prime h$, \ $g b \to t W^- h$, with the subsequent Higgs-boson decay to four leptons, $h \to 4 \ell$. 
Besides, in Ref.~\cite{CMS:2021nnc} measurement \cite{CMS:2020cga} of the channel $h \to \gamma \gamma$ was included in the analysis.
 
Below we take the following values of the couplings: (i) $\alpha = 1.05$, \ $\beta = 0.2$, (ii) $\alpha = 1.05$, \ $\beta = 0.44$ 
and (iii) $\alpha =1.05$, \ $\beta = 0.68$, choosing $\beta$ within the limits (\ref{eq:1029a}). 
It is sufficient to show asymmetries only for the positive $\beta$ since 
the couplings $\tilde{\kappa} (s)$ and $\tilde{\kappa}_z (s)$ in (\ref{eq:1022a}) are 
proportional to $\beta$, and therefore the asymmetries change sign for $\beta < 0$.

As is seen from Fig.~\ref{fig:A(s)_Eb-Ebbar}, for the value $\beta=0.68$ the asymmetry $A(s)$ can take value of about $ -10^{-3}$ at the electron (positron) energy $ E \sim 216 $ GeV. As for the difference of mean $b$ and $\bar{b}$ energies, 
it takes value of about $-$70 MeV at somewhat higher beam energy $E \sim 306$ GeV. 
The similar energy dependence of the asymmetries holds for other values of the parameter $\beta$. 
For $\beta = 0.44$, \ $ A(s) \approx - 0.65 \times 10^{-3}$ and 
$ \left\langle \varepsilon  \right\rangle - \left\langle \bar{\varepsilon} \right\rangle  \approx -46$ MeV, while for 
$\beta = 0.2$, \ $ A(s) \approx - 0.29 \times 10^{-3}$ and 
$ \left\langle \varepsilon  \right\rangle - \left\langle \bar{\varepsilon} \right\rangle  \approx -20$ MeV. Since the couplings $\tilde{\kappa} (s)$ and $\tilde{\kappa}_z (s)$ are linear in $\beta$, the energies at which 
the asymmetries take their maximal values do not depend on the new physics parameter $\beta$. 

\begin{figure}[tbh]
	\begin{center}
		\includegraphics[scale=0.61]{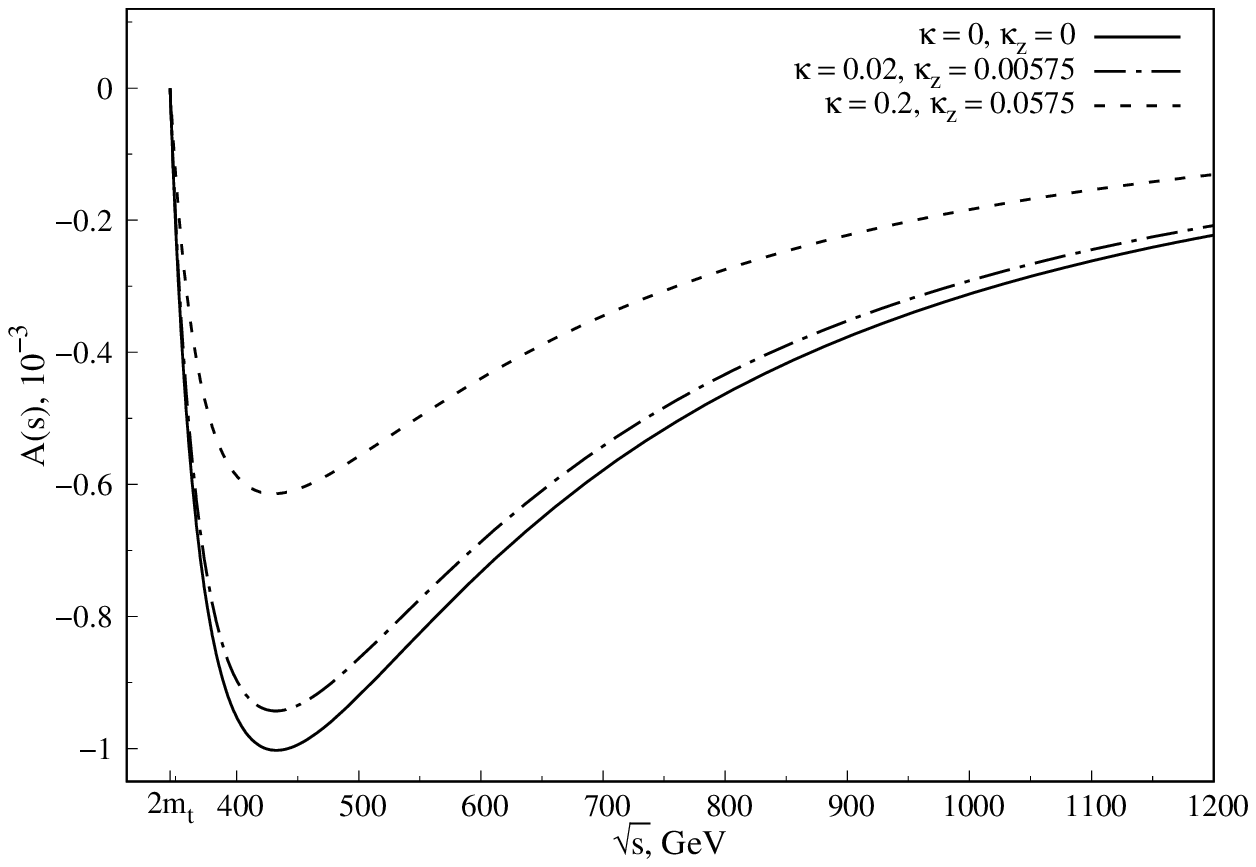}
		\includegraphics[scale=0.61]{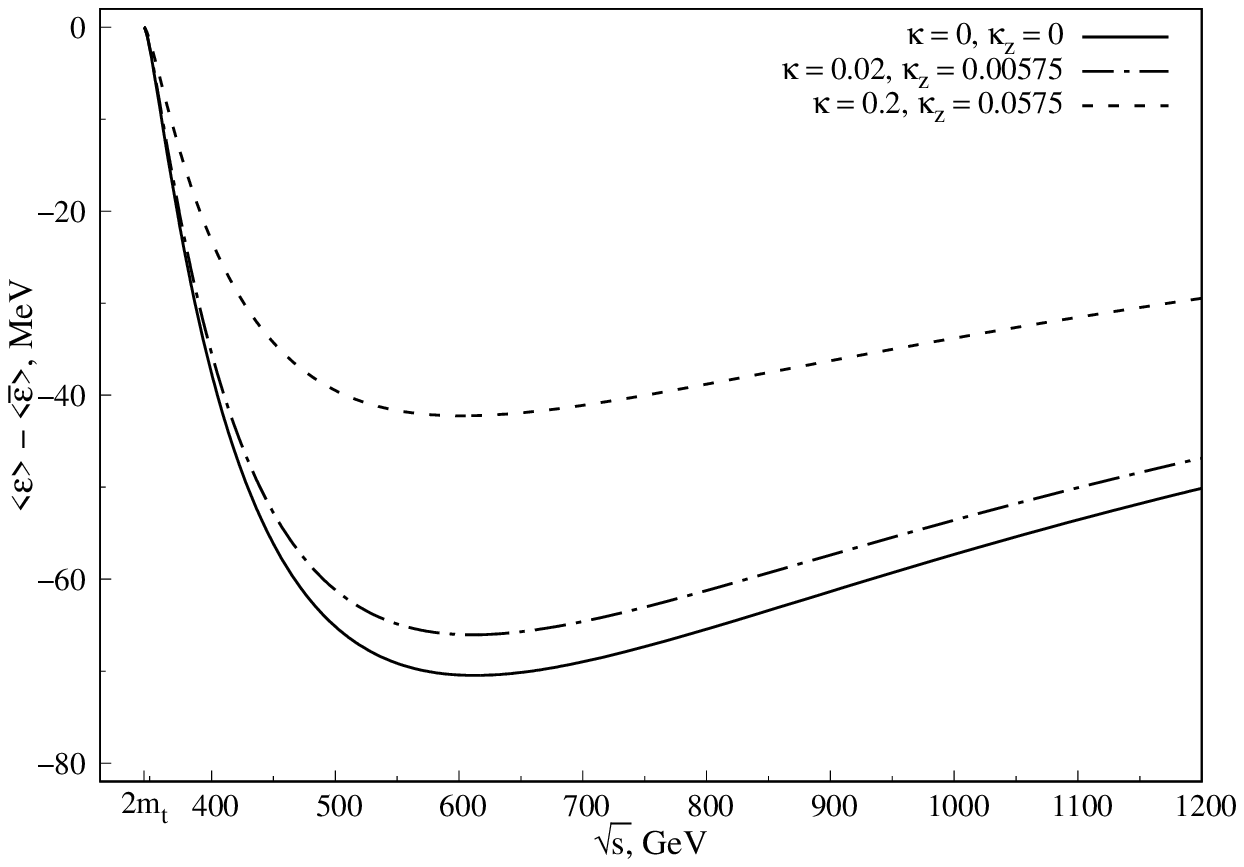}
	\end{center}
	\caption{The same asymmetries as in Fig.~\ref{fig:A(s)_Eb-Ebbar} for $\alpha =1.05$, $\beta = 0.68$, calculated with several values
	of $\kappa (s)$ and $\kappa_z (s)$ (the choice of these values is explained in the text).}
	\label{fig:A3(s)_E3b-E3bbar}
\end{figure}

Let us study dependence of the asymmetries on the $CP$-conserving couplings $\kappa$ and $\kappa_z$.
Results in Fig.~\ref{fig:A(s)_Eb-Ebbar} are shown for $\kappa = \kappa_z =0$ because contribution of these couplings to the $e^+ e^- \to t \bar{t}$ cross section $\sigma_0$ in the model of subsection~\ref{subsec:loop} 
is negligible. However, if $\kappa$ and $\kappa_z$ are not very small, they can alter sizably 
the $\sigma_0$ and thus the asymmetries. 

Indeed, in addition to the diagrams in Fig.~\ref{fig:5} there are other contributions to the $\gamma t t$ and $Z tt$ vertices. 
We mention radiative corrections (RC) to the vertices calculated in QCD to two loops~\cite{Bernreuther:2004ih, Bernreuther:2004th, Bernreuther:2005rw} 
and in EW theory to one loop~\cite{Czarnecki:1995sz, Hollik:1988ii, Bernabeu:1997je}.
Ref.~\cite{Bernreuther:2005gq}, in the two-loop order in QCD and in the lowest order in the EW couplings, 
predicts the following RC values at $s=0$
\begin{equation}
\kappa^{(RC)} = 0.020, \qquad \quad  \kappa_{z}^{(RC)} = 0.00575. 
\label{eq:1030}
\end{equation}
These values can give a rough estimate of RC at $s > 4 m_t^2$. Note that they are considerably greater than the values 
in Eqs.~(\ref{eq:1019}). 

In addition, there may exist effects of ``new physics'' (or BSM physics) which are not accounted for by the model in subsection~\ref{subsec:loop} 
and by RC in (\ref{eq:1030}). For a conservative estimate of the BSM effects we choose values of $\kappa^{(BSM)}$ and $\kappa_{z}^{(BSM)}$ ten times 
greater than the values in (\ref{eq:1030}). 
Note that these values are within the bounds on these couplings in  Eq.~(\ref{eq:05}).

The asymmetries (\ref{eq:1025}) and (\ref{eq:1027}) calculated with several values of $\kappa$ and $\kappa_z$ are shown in 
Fig.~\ref{fig:A3(s)_E3b-E3bbar}. It is seen that the asymmetries become sensitive to the couplings 
$\kappa$ and $\kappa_z$, if the latter have sizable values.
 

\section{\label{sec:conclusions} Conclusions}

We have studied the energy correlation of the bottom quark and antiquark from decays of the top quark and antiquark produced in the 
electron--positron annihilation. 
The cross section of the process $e^+ e^- \to b \, \bar{b} \, W^+ W^-$ as a function of the energies of $b$ and $\bar{b}$ quarks 
is derived and calculated. The main emphasis is put on investigation of $CP$-violation effects in the interaction of the photon and $Z$ boson with the
top quarks. The corresponding terms in the $\gamma t {t}$ and $Z t {t}$ vertices 
are determined by the couplings $\tilde{\kappa}(s)$ and $\tilde{\kappa}_z(s)$ 
related to EDM and WEDM, respectively.  In addition, the terms corresponding to AMDM and AWMDM, 
determined by the couplings $\kappa(s)$ and $\kappa_z(s)$, are included. 

The joint energy distribution of the bottom quark and antiquark $d^2 \sigma / dE_b \, dE_{\bar{b}}$ 
is calculated with these couplings. 
We considered observables which give access to the $CP$-violating terms. One of them is the
asymmetry in the energy distribution of $b$ and $\bar{b}$ quarks $A(s)$, the other one is the difference of mean 
energies of $b$ and $\bar{b}$ quarks $ \left\langle E_b \right\rangle - \left\langle E_{\bar{b}} \right\rangle$. 

In order to estimate magnitude of these observables we have calculated the couplings 
in one-loop model with intermediate Higgs boson. Interaction of this boson with the top quarks 
is assumed to include the scalar and pseudoscalar terms determined by the parameters $\alpha$ and $\beta$, respectively. 
In framework of this model the real and imaginary parts of $\tilde{\kappa}(s)$ and 
$\tilde{\kappa}_z(s)$ are calculated with the parameters $\alpha$ and $\beta$, values of which are taken from  
constraints reported by the CMS collaboration~\cite{CMS:2021nnc}. 
We studied dependence of the asymmetries on the $e^+ e^-$ energy up to $\sqrt{s}=1.2$ TeV. 
An interesting trend is observed: the asymmetries (modulo) have maximal values at certain $e^+ e^-$ energies, 
and the positions of these maxima are independent of the value of the parameter $\beta$.
Energy behavior of these observables can be of interest for future studies at CLIC at the next stages 
of its run, and for other future $e^+ e^-$ colliders. 

The influence of the $CP$-conserving couplings $\kappa(s)$ and $\kappa_z(s)$ on the asymmetries is also studied. 
The values of $\kappa(s)$ and $\kappa_z(s)$ are calculated in one-loop model mentioned above, or taken 
from the calculation~\cite{Bernreuther:2005gq} of the QCD and EW RC to the $\gamma t {t}$ and $Z t {t}$ vertices. 
It is shown that the asymmetries $A(s)$ and $\left\langle E_b \right\rangle - \left\langle E_{\bar{b}} \right\rangle$ 
become dependent on these couplings, if $\kappa(s)$ and $\kappa_z(s)$ are sufficiently big. This can be taken into 
consideration in future experiments which will aim at studying $CP$ violation in the $e^+ e^- \to t \bar{t}$ process. 

We should note that although the consideration in the paper has been performed for the unpolarized electron and positron, the present method can be extended to the case 
of the polarized electron and positron beams.

\section*{Acknowledgments}

This work was partially conducted in the scope of the IDEATE International Associated Laboratory (LIA).
The authors acknowledge partial support by the National Academy of Sciences of Ukraine 
via the programs ``Support for the development of priority areas of scientific research'' (6541230)
and ``Participation in the international projects in high energy and nuclear physics'' 
(project no. 0121U111693).


\bibliography{bib4-2}

\begin{thebibliography}{50}%
\makeatletter
\providecommand \@ifxundefined [1]{%
 \@ifx{#1\undefined}
}%
\providecommand \@ifnum [1]{%
 \ifnum #1\expandafter \@firstoftwo
 \else \expandafter \@secondoftwo
 \fi
}%
\providecommand \@ifx [1]{%
 \ifx #1\expandafter \@firstoftwo
 \else \expandafter \@secondoftwo
 \fi
}%
\providecommand \natexlab [1]{#1}%
\providecommand \enquote  [1]{``#1''}%
\providecommand \bibnamefont  [1]{#1}%
\providecommand \bibfnamefont [1]{#1}%
\providecommand \citenamefont [1]{#1}%
\providecommand \href@noop [0]{\@secondoftwo}%
\providecommand \href [0]{\begingroup \@sanitize@url \@href}%
\providecommand \@href[1]{\@@startlink{#1}\@@href}%
\providecommand \@@href[1]{\endgroup#1\@@endlink}%
\providecommand \@sanitize@url [0]{\catcode `\\12\catcode `\$12\catcode
  `\&12\catcode `\#12\catcode `\^12\catcode `\_12\catcode `\%12\relax}%
\providecommand \@@startlink[1]{}%
\providecommand \@@endlink[0]{}%
\providecommand \url  [0]{\begingroup\@sanitize@url \@url }%
\providecommand \@url [1]{\endgroup\@href {#1}{\urlprefix }}%
\providecommand \urlprefix  [0]{URL }%
\providecommand \Eprint [0]{\href }%
\providecommand \doibase [0]{https://doi.org/}%
\providecommand \selectlanguage [0]{\@gobble}%
\providecommand \bibinfo  [0]{\@secondoftwo}%
\providecommand \bibfield  [0]{\@secondoftwo}%
\providecommand \translation [1]{[#1]}%
\providecommand \BibitemOpen [0]{}%
\providecommand \bibitemStop [0]{}%
\providecommand \bibitemNoStop [0]{.\EOS\space}%
\providecommand \EOS [0]{\spacefactor3000\relax}%
\providecommand \BibitemShut  [1]{\csname bibitem#1\endcsname}%
\let\auto@bib@innerbib\@empty
\bibitem [{\citenamefont {Behnke}\ \emph {et~al.}(2013)\citenamefont {Behnke},
  \citenamefont {Brau}, \citenamefont {Foster}, \citenamefont {Fuster},
  \citenamefont {Harrison}, \citenamefont {Paterson}, \citenamefont {Peskin},
  \citenamefont {Stanitzki}, \citenamefont {Walker},\ and\ \citenamefont
  {Yamamoto}}]{Behnke:2013xla}%
  \BibitemOpen
  \bibfield  {author} {\bibinfo {author} {\bibfnamefont {T.}~\bibnamefont
  {Behnke}}, \bibinfo {author} {\bibfnamefont {J.~E.}\ \bibnamefont {Brau}},
  \bibinfo {author} {\bibfnamefont {B.}~\bibnamefont {Foster}}, \bibinfo
  {author} {\bibfnamefont {J.}~\bibnamefont {Fuster}}, \bibinfo {author}
  {\bibfnamefont {M.}~\bibnamefont {Harrison}}, \bibinfo {author}
  {\bibfnamefont {J.~M.}\ \bibnamefont {Paterson}}, \bibinfo {author}
  {\bibfnamefont {M.}~\bibnamefont {Peskin}}, \bibinfo {author} {\bibfnamefont
  {M.}~\bibnamefont {Stanitzki}}, \bibinfo {author} {\bibfnamefont
  {N.}~\bibnamefont {Walker}},\ and\ \bibinfo {author} {\bibfnamefont
  {H.}~\bibnamefont {Yamamoto}},\ }\href {https://doi.org/10.2172/1347945}
  {\emph {\bibinfo {title} {The International Linear Collider Technical Design
  Report - Volume 1: Executive Summary}}},\ \bibinfo {type} {Tech. Rep.}\
  (\bibinfo  {institution} {CERN},\ \bibinfo {address} {Geneva},\ \bibinfo
  {year} {2013})\ \Eprint {https://arxiv.org/abs/1306.6327} {arXiv:1306.6327
  [physics.acc-ph]} \BibitemShut {NoStop}%
\bibitem [{\citenamefont {Baer}\ \emph {et~al.}(2013)\citenamefont {Baer} \emph
  {et~al.}}]{Baer:2013cma}%
  \BibitemOpen
  \bibfield  {author} {\bibinfo {author} {\bibfnamefont {H.}~\bibnamefont
  {Baer}} \emph {et~al.},\ }\href@noop {} {\emph {\bibinfo {title} {{The
  International Linear Collider Technical Design Report - Volume 2:
  Physics}}}},\ \bibinfo {type} {Tech. Rep.}\ (\bibinfo  {institution} {CERN},\
  \bibinfo {address} {Geneva},\ \bibinfo {year} {2013})\ \Eprint
  {https://arxiv.org/abs/1306.6352} {arXiv:1306.6352 [hep-ph]} \BibitemShut
  {NoStop}%
\bibitem [{\citenamefont {Yamamoto}(2021)}]{Yamamoto:2021kig}%
  \BibitemOpen
  \bibfield  {author} {\bibinfo {author} {\bibfnamefont {H.}~\bibnamefont
  {Yamamoto}},\ }\bibfield  {title} {\bibinfo {title} {{The International
  Linear Collider Project \textemdash{} Its Physics and Status}},\ }\href
  {https://doi.org/10.3390/sym13040674} {\bibfield  {journal} {\bibinfo
  {journal} {Symmetry}\ }\textbf {\bibinfo {volume} {13}},\ \bibinfo {pages}
  {674} (\bibinfo {year} {2021})}\BibitemShut {NoStop}%
\bibitem [{\citenamefont {Aicheler}\ \emph {et~al.}(2012)\citenamefont
  {Aicheler}, \citenamefont {Burrows}, \citenamefont {Draper}, \citenamefont
  {Garvey}, \citenamefont {Lebrun}, \citenamefont {Peach}, \citenamefont
  {Phinney}, \citenamefont {Schmickler}, \citenamefont {Schulte},\ and\
  \citenamefont {Toge}}]{Aicheler:2012bya}%
  \BibitemOpen
  \bibfield  {author} {\bibinfo {author} {\bibfnamefont {M.}~\bibnamefont
  {Aicheler}}, \bibinfo {author} {\bibfnamefont {P.}~\bibnamefont {Burrows}},
  \bibinfo {author} {\bibfnamefont {M.}~\bibnamefont {Draper}}, \bibinfo
  {author} {\bibfnamefont {T.}~\bibnamefont {Garvey}}, \bibinfo {author}
  {\bibfnamefont {P.}~\bibnamefont {Lebrun}}, \bibinfo {author} {\bibfnamefont
  {K.}~\bibnamefont {Peach}}, \bibinfo {author} {\bibfnamefont
  {N.}~\bibnamefont {Phinney}}, \bibinfo {author} {\bibfnamefont
  {H.}~\bibnamefont {Schmickler}}, \bibinfo {author} {\bibfnamefont
  {D.}~\bibnamefont {Schulte}},\ and\ \bibinfo {author} {\bibfnamefont
  {N.}~\bibnamefont {Toge}},\ }\href {https://doi.org/10.5170/CERN-2012-007}
  {\emph {\bibinfo {title} {{A Multi-TeV Linear Collider Based on CLIC
  Technology}: {CLIC Conceptual Design Report}}}},\ CERN Yellow Reports:
  Monographs\ (\bibinfo  {publisher} {CERN},\ \bibinfo {address} {Geneva},\
  \bibinfo {year} {2012})\BibitemShut {NoStop}%
\bibitem [{\citenamefont {de~Blas}\ \emph {et~al.}(2018)\citenamefont {de~Blas}
  \emph {et~al.}}]{deBlas:2018mhx}%
  \BibitemOpen
  \bibfield  {author} {\bibinfo {author} {\bibfnamefont {J.}~\bibnamefont
  {de~Blas}} \emph {et~al.},\ }\href {https://doi.org/10.23731/CYRM-2018-003}
  {\emph {\bibinfo {title} {{The CLIC Potential for New Physics}}}},\ CERN
  Yellow Reports: Monographs\ (\bibinfo  {publisher} {CERN},\ \bibinfo
  {address} {Geneva},\ \bibinfo {year} {2018})\ \Eprint
  {https://arxiv.org/abs/1812.02093} {arXiv:1812.02093 [hep-ph]} \BibitemShut
  {NoStop}%
\bibitem [{\citenamefont {Zarnecki}(2019)}]{Zarnecki:2019vrn}%
  \BibitemOpen
  \bibfield  {author} {\bibinfo {author} {\bibfnamefont {A.~F.}\ \bibnamefont
  {Zarnecki}} (\bibinfo {collaboration} {CLICdp}),\ }\bibfield  {title}
  {\bibinfo {title} {{News on the CLIC physics potential}},\ }\href
  {https://doi.org/10.22323/1.360.0008} {\bibfield  {journal} {\bibinfo
  {journal} {PoS}\ }\textbf {\bibinfo {volume} {ALPS2019}},\ \bibinfo {pages}
  {008} (\bibinfo {year} {2019})},\ \Eprint {https://arxiv.org/abs/1908.04671}
  {arXiv:1908.04671 [hep-ex]} \BibitemShut {NoStop}%
\bibitem [{\citenamefont {Zarnecki}(2020)}]{Zarnecki:2020ics}%
  \BibitemOpen
  \bibfield  {author} {\bibinfo {author} {\bibfnamefont {A.~F.}\ \bibnamefont
  {Zarnecki}} (\bibinfo {collaboration} {CLICdp, ILD concept group}),\
  }\bibfield  {title} {\bibinfo {title} {{On the physics potential of ILC and
  CLIC}},\ }\href {https://doi.org/10.22323/1.376.0037} {\bibfield  {journal}
  {\bibinfo  {journal} {PoS}\ }\textbf {\bibinfo {volume} {CORFU2019}},\
  \bibinfo {pages} {037} (\bibinfo {year} {2020})},\ \Eprint
  {https://arxiv.org/abs/2004.14628} {arXiv:2004.14628 [hep-ph]} \BibitemShut
  {NoStop}%
\bibitem [{\citenamefont {Kemppinen}\ \emph {et~al.}(2021)\citenamefont
  {Kemppinen}, \citenamefont {Rude}, \citenamefont {Mainaud~Durand},\ and\
  \citenamefont {Mattila}}]{Kemppinen:2021urj}%
  \BibitemOpen
  \bibfield  {author} {\bibinfo {author} {\bibfnamefont {J.}~\bibnamefont
  {Kemppinen}}, \bibinfo {author} {\bibfnamefont {V.}~\bibnamefont {Rude}},
  \bibinfo {author} {\bibfnamefont {H.}~\bibnamefont {Mainaud~Durand}},\ and\
  \bibinfo {author} {\bibfnamefont {J.}~\bibnamefont {Mattila}},\ }\bibfield
  {title} {\bibinfo {title} {{CLIC pre-alignment \textemdash{} status and
  remaining challenges}},\ }\href {https://doi.org/10.1088/1361-6501/ac09b3}
  {\bibfield  {journal} {\bibinfo  {journal} {Measur. Sci. Tech.}\ }\textbf
  {\bibinfo {volume} {32}},\ \bibinfo {pages} {115015} (\bibinfo {year}
  {2021})}\BibitemShut {NoStop}%
\bibitem [{\citenamefont {Bambade}\ \emph {et~al.}(2019)\citenamefont {Bambade}
  \emph {et~al.}}]{Bambade:2019fyw}%
  \BibitemOpen
  \bibfield  {author} {\bibinfo {author} {\bibfnamefont {P.}~\bibnamefont
  {Bambade}} \emph {et~al.},\ }\bibfield  {title} {\bibinfo {title} {The
  international linear collider: A global project},\ }\href@noop {} {\
  (\bibinfo {year} {2019})},\ \Eprint {https://arxiv.org/abs/1903.01629}
  {arXiv:1903.01629 [hep-ex]} \BibitemShut {NoStop}%
\bibitem [{\citenamefont {Charles}\ \emph {et~al.}(2018)\citenamefont {Charles}
  \emph {et~al.}}]{CLICdp:2018cto}%
  \BibitemOpen
  \bibfield  {author} {\bibinfo {author} {\bibfnamefont {T.~K.}\ \bibnamefont
  {Charles}} \emph {et~al.} (\bibinfo {collaboration} {CLIC, CLICdp}),\ }\href
  {https://doi.org/10.23731/CYRM-2018-002} {\emph {\bibinfo {title} {{The
  Compact Linear Collider (CLIC) - 2018 Summary Report}}}},\ \bibinfo {series}
  {CERN Yellow Reports: Monographs}, Vol.~\bibinfo {volume} {2}\ (\bibinfo
  {publisher} {CERN},\ \bibinfo {year} {2018})\BibitemShut {NoStop}%
\bibitem [{\citenamefont {Roloff}\ \emph {et~al.}(2018)\citenamefont {Roloff},
  \citenamefont {Franceschini}, \citenamefont {Schnoor},\ and\ \citenamefont
  {Wulzer}}]{Roloff:2018dqu}%
  \BibitemOpen
  \bibfield  {author} {\bibinfo {author} {\bibfnamefont {P.}~\bibnamefont
  {Roloff}}, \bibinfo {author} {\bibfnamefont {R.}~\bibnamefont
  {Franceschini}}, \bibinfo {author} {\bibfnamefont {U.}~\bibnamefont
  {Schnoor}},\ and\ \bibinfo {author} {\bibfnamefont {A.}~\bibnamefont
  {Wulzer}} (\bibinfo {collaboration} {CLIC, CLICdp}),\ }\href@noop {} {\emph
  {\bibinfo {title} {{The Compact Linear e$^+$e$^-$ Collider (CLIC): Physics
  Potential}}}},\ \bibinfo {type} {Tech. Rep.}\ (\bibinfo  {institution}
  {CERN},\ \bibinfo {year} {2018})\ \Eprint {https://arxiv.org/abs/1812.07986}
  {arXiv:1812.07986 [hep-ex]} \BibitemShut {NoStop}%
\bibitem [{\citenamefont {Abada}\ \emph
  {et~al.}(2019{\natexlab{a}})\citenamefont {Abada} \emph
  {et~al.}}]{FCC:2018byv}%
  \BibitemOpen
  \bibfield  {author} {\bibinfo {author} {\bibfnamefont {A.}~\bibnamefont
  {Abada}} \emph {et~al.} (\bibinfo {collaboration} {FCC}),\ }\bibfield
  {title} {\bibinfo {title} {{FCC Physics Opportunities}: {Future Circular
  Collider Conceptual Design Report Volume 1}},\ }\href
  {https://doi.org/10.1140/epjc/s10052-019-6904-3} {\bibfield  {journal}
  {\bibinfo  {journal} {Eur. Phys. J. C}\ }\textbf {\bibinfo {volume} {79}},\
  \bibinfo {pages} {474} (\bibinfo {year} {2019}{\natexlab{a}})}\BibitemShut
  {NoStop}%
\bibitem [{\citenamefont {Abada}\ \emph
  {et~al.}(2019{\natexlab{b}})\citenamefont {Abada} \emph
  {et~al.}}]{FCC:2018evy}%
  \BibitemOpen
  \bibfield  {author} {\bibinfo {author} {\bibfnamefont {A.}~\bibnamefont
  {Abada}} \emph {et~al.} (\bibinfo {collaboration} {FCC}),\ }\bibfield
  {title} {\bibinfo {title} {{FCC-ee: The Lepton Collider}: {Future Circular
  Collider Conceptual Design Report Volume 2}},\ }\href
  {https://doi.org/10.1140/epjst/e2019-900045-4} {\bibfield  {journal}
  {\bibinfo  {journal} {Eur. Phys. J. ST}\ }\textbf {\bibinfo {volume} {228}},\
  \bibinfo {pages} {261} (\bibinfo {year} {2019}{\natexlab{b}})}\BibitemShut
  {NoStop}%
\bibitem [{\citenamefont {Arens}\ and\ \citenamefont
  {Sehgal}(1994)}]{Arens:1994jp}%
  \BibitemOpen
  \bibfield  {author} {\bibinfo {author} {\bibfnamefont {T.}~\bibnamefont
  {Arens}}\ and\ \bibinfo {author} {\bibfnamefont {L.~M.}\ \bibnamefont
  {Sehgal}},\ }\bibfield  {title} {\bibinfo {title} {{Energy correlation and
  asymmetry of secondary leptons in $e^{+} e^{-} \to t \bar{t}$}},\ }\href
  {https://doi.org/10.1103/PhysRevD.50.4372} {\bibfield  {journal} {\bibinfo
  {journal} {Phys. Rev. D}\ }\textbf {\bibinfo {volume} {50}},\ \bibinfo
  {pages} {4372} (\bibinfo {year} {1994})}\BibitemShut {NoStop}%
\bibitem [{\citenamefont {Arens}\ and\ \citenamefont
  {Sehgal}(1993)}]{Arens:1992wh}%
  \BibitemOpen
  \bibfield  {author} {\bibinfo {author} {\bibfnamefont {T.}~\bibnamefont
  {Arens}}\ and\ \bibinfo {author} {\bibfnamefont {L.~M.}\ \bibnamefont
  {Sehgal}},\ }\bibfield  {title} {\bibinfo {title} {{Secondary leptons as
  probes of top quark polarization in $e^{+} e^{-} \to t \bar{t}$}},\ }\href
  {https://doi.org/10.1016/0550-3213(93)90236-I} {\bibfield  {journal}
  {\bibinfo  {journal} {Nucl. Phys. B}\ }\textbf {\bibinfo {volume} {393}},\
  \bibinfo {pages} {46} (\bibinfo {year} {1993})}\BibitemShut {NoStop}%
\bibitem [{\citenamefont {Grzadkowski}\ and\ \citenamefont
  {Hioki}(1997)}]{Grzadkowski:1996kn}%
  \BibitemOpen
  \bibfield  {author} {\bibinfo {author} {\bibfnamefont {B.}~\bibnamefont
  {Grzadkowski}}\ and\ \bibinfo {author} {\bibfnamefont {Z.}~\bibnamefont
  {Hioki}},\ }\bibfield  {title} {\bibinfo {title} {{Energy spectrum of
  secondary leptons in $e^{+} e^{-} \to t \bar{t}$: Nonstandard effects and CP
  violation}},\ }\href {https://doi.org/10.1016/S0550-3213(96)00602-5}
  {\bibfield  {journal} {\bibinfo  {journal} {Nucl. Phys. B}\ }\textbf
  {\bibinfo {volume} {484}},\ \bibinfo {pages} {17} (\bibinfo {year} {1997})},\
  \Eprint {https://arxiv.org/abs/hep-ph/9604301} {arXiv:hep-ph/9604301}
  \BibitemShut {NoStop}%
\bibitem [{\citenamefont {Christova}(1999)}]{Christova:1998et}%
  \BibitemOpen
  \bibfield  {author} {\bibinfo {author} {\bibfnamefont {E.}~\bibnamefont
  {Christova}},\ }\bibfield  {title} {\bibinfo {title} {{CP violating
  asymmetries of $b$ quarks and leptons in $e^{+} e^{-} \to t \bar{t}$ and
  supersymmetry}},\ }\href {https://doi.org/10.1142/S0217751X99002451}
  {\bibfield  {journal} {\bibinfo  {journal} {Int. J. Mod. Phys. A}\ }\textbf
  {\bibinfo {volume} {14}},\ \bibinfo {pages} {1} (\bibinfo {year} {1999})},\
  \Eprint {https://arxiv.org/abs/hep-ph/9809290} {arXiv:hep-ph/9809290}
  \BibitemShut {NoStop}%
\bibitem [{\citenamefont {Bartl}\ \emph {et~al.}(1999)\citenamefont {Bartl},
  \citenamefont {Christova}, \citenamefont {Gajdosik},\ and\ \citenamefont
  {Majerotto}}]{Bartl:1998nn}%
  \BibitemOpen
  \bibfield  {author} {\bibinfo {author} {\bibfnamefont {A.}~\bibnamefont
  {Bartl}}, \bibinfo {author} {\bibfnamefont {E.}~\bibnamefont {Christova}},
  \bibinfo {author} {\bibfnamefont {T.}~\bibnamefont {Gajdosik}},\ and\
  \bibinfo {author} {\bibfnamefont {W.}~\bibnamefont {Majerotto}},\ }\bibfield
  {title} {\bibinfo {title} {{CP violating energy asymmetries of $b$ and
  $\bar{b}$ quarks in $e^{+} e^{-} \to t \bar{t}$}},\ }\href
  {https://doi.org/10.1103/PhysRevD.59.077503} {\bibfield  {journal} {\bibinfo
  {journal} {Phys. Rev. D}\ }\textbf {\bibinfo {volume} {59}},\ \bibinfo
  {pages} {077503} (\bibinfo {year} {1999})},\ \Eprint
  {https://arxiv.org/abs/hep-ph/9803426} {arXiv:hep-ph/9803426} \BibitemShut
  {NoStop}%
\bibitem [{\citenamefont {Kawasaki}\ \emph {et~al.}(1973)\citenamefont
  {Kawasaki}, \citenamefont {Shirafuji},\ and\ \citenamefont
  {Tsai}}]{Kawasaki:1973hf}%
  \BibitemOpen
  \bibfield  {author} {\bibinfo {author} {\bibfnamefont {S.}~\bibnamefont
  {Kawasaki}}, \bibinfo {author} {\bibfnamefont {T.}~\bibnamefont
  {Shirafuji}},\ and\ \bibinfo {author} {\bibfnamefont {S.~Y.}\ \bibnamefont
  {Tsai}},\ }\bibfield  {title} {\bibinfo {title} {{Productions and decays of
  short-lived particles in $e^{+} e^{-}$ colliding beam experiments}},\ }\href
  {https://doi.org/10.1143/PTP.49.1656} {\bibfield  {journal} {\bibinfo
  {journal} {Prog. Theor. Phys.}\ }\textbf {\bibinfo {volume} {49}},\ \bibinfo
  {pages} {1656} (\bibinfo {year} {1973})}\BibitemShut {NoStop}%
\bibitem [{\citenamefont {Hollik}\ \emph {et~al.}(1999)\citenamefont {Hollik},
  \citenamefont {Illana}, \citenamefont {Rigolin}, \citenamefont
  {Schappacher},\ and\ \citenamefont {Stockinger}}]{Hollik:1998vz}%
  \BibitemOpen
  \bibfield  {author} {\bibinfo {author} {\bibfnamefont {W.}~\bibnamefont
  {Hollik}}, \bibinfo {author} {\bibfnamefont {J.~I.}\ \bibnamefont {Illana}},
  \bibinfo {author} {\bibfnamefont {S.}~\bibnamefont {Rigolin}}, \bibinfo
  {author} {\bibfnamefont {C.}~\bibnamefont {Schappacher}},\ and\ \bibinfo
  {author} {\bibfnamefont {D.}~\bibnamefont {Stockinger}},\ }\bibfield  {title}
  {\bibinfo {title} {{Top dipole form-factors and loop induced CP violation in
  supersymmetry}},\ }\href {https://doi.org/10.1016/S0550-3213(99)00396-X}
  {\bibfield  {journal} {\bibinfo  {journal} {Nucl. Phys. B}\ }\textbf
  {\bibinfo {volume} {551}},\ \bibinfo {pages} {3} (\bibinfo {year} {1999})},\
  \bibinfo {note} {[Erratum: Nucl. Phys. B 557, 407--409 (1999)]},\ \Eprint
  {https://arxiv.org/abs/hep-ph/9812298} {arXiv:hep-ph/9812298} \BibitemShut
  {NoStop}%
\bibitem [{\citenamefont {Atwood}\ \emph {et~al.}(2001)\citenamefont {Atwood},
  \citenamefont {Bar-Shalom}, \citenamefont {Eilam},\ and\ \citenamefont
  {Soni}}]{Atwood:2000tu}%
  \BibitemOpen
  \bibfield  {author} {\bibinfo {author} {\bibfnamefont {D.}~\bibnamefont
  {Atwood}}, \bibinfo {author} {\bibfnamefont {S.}~\bibnamefont {Bar-Shalom}},
  \bibinfo {author} {\bibfnamefont {G.}~\bibnamefont {Eilam}},\ and\ \bibinfo
  {author} {\bibfnamefont {A.}~\bibnamefont {Soni}},\ }\bibfield  {title}
  {\bibinfo {title} {{CP violation in top physics}},\ }\href
  {https://doi.org/10.1016/S0370-1573(00)00112-5} {\bibfield  {journal}
  {\bibinfo  {journal} {Phys. Rept.}\ }\textbf {\bibinfo {volume} {347}},\
  \bibinfo {pages} {1} (\bibinfo {year} {2001})},\ \Eprint
  {https://arxiv.org/abs/hep-ph/0006032} {arXiv:hep-ph/0006032} \BibitemShut
  {NoStop}%
\bibitem [{\citenamefont {Zhang}(2020)}]{Zhang:2020jxw}%
  \BibitemOpen
  \bibfield  {author} {\bibinfo {author} {\bibfnamefont {Y.}~\bibnamefont
  {Zhang}},\ }\emph {\bibinfo {title} {{Prospects for Precision Measurements of
  the Top-Yukawa Coupling and CP Violation in $t\bar{t}H$ Production at the
  CLIC $e^+e^-$ Collider}}},\ \href@noop {} {Ph.D. thesis},\ \bibinfo  {school}
  {Edinburgh U.} (\bibinfo {year} {2020})\BibitemShut {NoStop}%
\bibitem [{\citenamefont {Faroughy}\ \emph {et~al.}(2021)\citenamefont
  {Faroughy}, \citenamefont {Bortolato}, \citenamefont {Kamenik}, \citenamefont
  {Ko\v{s}nik},\ and\ \citenamefont {Smolkovi\v{c}}}]{Faroughy:2021sxk}%
  \BibitemOpen
  \bibfield  {author} {\bibinfo {author} {\bibfnamefont {D.~A.}\ \bibnamefont
  {Faroughy}}, \bibinfo {author} {\bibfnamefont {B.}~\bibnamefont {Bortolato}},
  \bibinfo {author} {\bibfnamefont {J.~F.}\ \bibnamefont {Kamenik}}, \bibinfo
  {author} {\bibfnamefont {N.}~\bibnamefont {Ko\v{s}nik}},\ and\ \bibinfo
  {author} {\bibfnamefont {A.}~\bibnamefont {Smolkovi\v{c}}},\ }\bibfield
  {title} {\bibinfo {title} {{Optimized Probes of the $CP$ Nature of the Top
  Quark Yukawa Coupling at Hadron Colliders}},\ }\href
  {https://doi.org/10.3390/sym13071129} {\bibfield  {journal} {\bibinfo
  {journal} {Symmetry}\ }\textbf {\bibinfo {volume} {13}},\ \bibinfo {pages}
  {1129} (\bibinfo {year} {2021})}\BibitemShut {NoStop}%
\bibitem [{\citenamefont {Bernreuther}\ \emph
  {et~al.}(1992{\natexlab{a}})\citenamefont {Bernreuther}, \citenamefont
  {Schroder},\ and\ \citenamefont {Pham}}]{Bernreuther:1992dz}%
  \BibitemOpen
  \bibfield  {author} {\bibinfo {author} {\bibfnamefont {W.}~\bibnamefont
  {Bernreuther}}, \bibinfo {author} {\bibfnamefont {T.}~\bibnamefont
  {Schroder}},\ and\ \bibinfo {author} {\bibfnamefont {T.~N.}\ \bibnamefont
  {Pham}},\ }\bibfield  {title} {\bibinfo {title} {{CP violating dipole
  form-factors in $e^{+} e^{-} \to t \bar{t}$}},\ }\href
  {https://doi.org/10.1016/0370-2693(92)90410-6} {\bibfield  {journal}
  {\bibinfo  {journal} {Phys. Lett. B}\ }\textbf {\bibinfo {volume} {279}},\
  \bibinfo {pages} {389} (\bibinfo {year} {1992}{\natexlab{a}})}\BibitemShut
  {NoStop}%
\bibitem [{\citenamefont {Chang}\ \emph {et~al.}(1993)\citenamefont {Chang},
  \citenamefont {Keung},\ and\ \citenamefont {Phillips}}]{Chang:1993fu}%
  \BibitemOpen
  \bibfield  {author} {\bibinfo {author} {\bibfnamefont {D.}~\bibnamefont
  {Chang}}, \bibinfo {author} {\bibfnamefont {W.-Y.}\ \bibnamefont {Keung}},\
  and\ \bibinfo {author} {\bibfnamefont {I.}~\bibnamefont {Phillips}},\
  }\bibfield  {title} {\bibinfo {title} {{CP violation in top pair production
  at an $e^+ e^-$ collider}},\ }\href
  {https://doi.org/10.1016/0550-3213(93)90536-X} {\bibfield  {journal}
  {\bibinfo  {journal} {Nucl. Phys. B}\ }\textbf {\bibinfo {volume} {408}},\
  \bibinfo {pages} {286} (\bibinfo {year} {1993})},\ \bibinfo {note} {[Erratum:
  Nucl. Phys. B 429, 255 (1994)]},\ \Eprint
  {https://arxiv.org/abs/hep-ph/9301259} {arXiv:hep-ph/9301259} \BibitemShut
  {NoStop}%
\bibitem [{\citenamefont {Sirunyan}\ \emph {et~al.}(2021)\citenamefont
  {Sirunyan} \emph {et~al.}}]{CMS:2021nnc}%
  \BibitemOpen
  \bibfield  {author} {\bibinfo {author} {\bibfnamefont {A.~M.}\ \bibnamefont
  {Sirunyan}} \emph {et~al.} (\bibinfo {collaboration} {CMS Collaboration}),\
  }\bibfield  {title} {\bibinfo {title} {{Constraints on anomalous Higgs boson
  couplings to vector bosons and fermions in its production and decay using the
  four-lepton final state}},\ }\href
  {https://doi.org/10.1103/PhysRevD.104.052004} {\bibfield  {journal} {\bibinfo
   {journal} {Phys. Rev. D}\ }\textbf {\bibinfo {volume} {104}},\ \bibinfo
  {pages} {052004} (\bibinfo {year} {2021})},\ \Eprint
  {https://arxiv.org/abs/2104.12152} {arXiv:2104.12152 [hep-ex]} \BibitemShut
  {NoStop}%
\bibitem [{\citenamefont {Bernreuther}\ \emph
  {et~al.}(1992{\natexlab{b}})\citenamefont {Bernreuther}, \citenamefont
  {Nachtmann}, \citenamefont {Overmann},\ and\ \citenamefont
  {Schr\"oder}}]{Bernreuther:1992be}%
  \BibitemOpen
  \bibfield  {author} {\bibinfo {author} {\bibfnamefont {W.}~\bibnamefont
  {Bernreuther}}, \bibinfo {author} {\bibfnamefont {O.}~\bibnamefont
  {Nachtmann}}, \bibinfo {author} {\bibfnamefont {P.}~\bibnamefont
  {Overmann}},\ and\ \bibinfo {author} {\bibfnamefont {T.}~\bibnamefont
  {Schr\"oder}},\ }\bibfield  {title} {\bibinfo {title} {{Angular correlations
  and distributions for searches of CP violation in top quark production and
  decay}},\ }\href {https://doi.org/10.1016/0550-3213(92)90545-M} {\bibfield
  {journal} {\bibinfo  {journal} {Nucl. Phys. B}\ }\textbf {\bibinfo {volume}
  {388}},\ \bibinfo {pages} {53} (\bibinfo {year} {1992}{\natexlab{b}})},\
  \bibinfo {note} {[Erratum: Nucl. Phys. B 406, 516--516 (1993)]}\BibitemShut
  {NoStop}%
\bibitem [{\citenamefont {Bernreuther}\ \emph {et~al.}(1993)\citenamefont
  {Bernreuther}, \citenamefont {Nachtmann},\ and\ \citenamefont
  {Overmann}}]{Bernreuther:1993nd}%
  \BibitemOpen
  \bibfield  {author} {\bibinfo {author} {\bibfnamefont {W.}~\bibnamefont
  {Bernreuther}}, \bibinfo {author} {\bibfnamefont {O.}~\bibnamefont
  {Nachtmann}},\ and\ \bibinfo {author} {\bibfnamefont {P.}~\bibnamefont
  {Overmann}},\ }\bibfield  {title} {\bibinfo {title} {{The CP violating
  electric and weak dipole moments of the tau lepton from threshold to
  500-GeV}},\ }\href {https://doi.org/10.1103/PhysRevD.48.78} {\bibfield
  {journal} {\bibinfo  {journal} {Phys. Rev. D}\ }\textbf {\bibinfo {volume}
  {48}},\ \bibinfo {pages} {78} (\bibinfo {year} {1993})}\BibitemShut {NoStop}%
\bibitem [{\citenamefont {Truten}\ and\ \citenamefont
  {Korchin}(2021)}]{Truten:2021nkj}%
  \BibitemOpen
  \bibfield  {author} {\bibinfo {author} {\bibfnamefont {I.~V.}\ \bibnamefont
  {Truten}}\ and\ \bibinfo {author} {\bibfnamefont {A.~Y.}\ \bibnamefont
  {Korchin}},\ }\bibfield  {title} {\bibinfo {title} {{Energy and angular
  distributions of the bottom quark in the electron--positron annihilation
  $e^+e^-\to b \, W^+ \, \bar{t}$}},\ }\href
  {https://doi.org/10.1142/S0217751X21500135} {\bibfield  {journal} {\bibinfo
  {journal} {Int. J. Mod. Phys. A}\ }\textbf {\bibinfo {volume} {36}},\
  \bibinfo {pages} {2150013} (\bibinfo {year} {2021})},\ \Eprint
  {https://arxiv.org/abs/2009.00301} {arXiv:2009.00301 [hep-ph]} \BibitemShut
  {NoStop}%
\bibitem [{\citenamefont {Truten}\ and\ \citenamefont
  {Korchin}(2019)}]{Truten:2019sqb}%
  \BibitemOpen
  \bibfield  {author} {\bibinfo {author} {\bibfnamefont {I.~V.}\ \bibnamefont
  {Truten}}\ and\ \bibinfo {author} {\bibfnamefont {A.~Y.}\ \bibnamefont
  {Korchin}},\ }\bibfield  {title} {\bibinfo {title} {{The top-quark
  polarization beyond the Standard Model in electron\textendash{}positron
  annihilation}},\ }\href {https://doi.org/10.1142/S0217751X19500672}
  {\bibfield  {journal} {\bibinfo  {journal} {Int. J. Mod. Phys. A}\ }\textbf
  {\bibinfo {volume} {34}},\ \bibinfo {pages} {1950067} (\bibinfo {year}
  {2019})},\ \Eprint {https://arxiv.org/abs/1902.09911} {arXiv:1902.09911
  [hep-ph]} \BibitemShut {NoStop}%
\bibitem [{\citenamefont {Bouzas}\ and\ \citenamefont
  {Larios}(2021)}]{Bouzas:2021gwx}%
  \BibitemOpen
  \bibfield  {author} {\bibinfo {author} {\bibfnamefont {A.~O.}\ \bibnamefont
  {Bouzas}}\ and\ \bibinfo {author} {\bibfnamefont {F.}~\bibnamefont
  {Larios}},\ }\bibfield  {title} {\bibinfo {title} {{Top quark effective
  couplings from top-pair tagged photoproduction in $pe^-$ collisions}},\
  }\href@noop {} {\  (\bibinfo {year} {2021})},\ \Eprint
  {https://arxiv.org/abs/2111.04723} {arXiv:2111.04723 [hep-ph]} \BibitemShut
  {NoStop}%
\bibitem [{\citenamefont {Grzadkowski}\ \emph {et~al.}(2010)\citenamefont
  {Grzadkowski}, \citenamefont {Iskrzynski}, \citenamefont {Misiak},\ and\
  \citenamefont {Rosiek}}]{Grzadkowski:2010es}%
  \BibitemOpen
  \bibfield  {author} {\bibinfo {author} {\bibfnamefont {B.}~\bibnamefont
  {Grzadkowski}}, \bibinfo {author} {\bibfnamefont {M.}~\bibnamefont
  {Iskrzynski}}, \bibinfo {author} {\bibfnamefont {M.}~\bibnamefont {Misiak}},\
  and\ \bibinfo {author} {\bibfnamefont {J.}~\bibnamefont {Rosiek}},\
  }\bibfield  {title} {\bibinfo {title} {{Dimension-Six Terms in the Standard
  Model Lagrangian}},\ }\href {https://doi.org/10.1007/JHEP10(2010)085}
  {\bibfield  {journal} {\bibinfo  {journal} {JHEP}\ }\textbf {\bibinfo
  {volume} {10}},\ \bibinfo {pages} {085}},\ \Eprint
  {https://arxiv.org/abs/1008.4884} {arXiv:1008.4884 [hep-ph]} \BibitemShut
  {NoStop}%
\bibitem [{\citenamefont {Zyla}\ \emph {et~al.}(2020)\citenamefont {Zyla} \emph
  {et~al.}}]{Zyla:2020zbs}%
  \BibitemOpen
  \bibfield  {author} {\bibinfo {author} {\bibfnamefont {P.}~\bibnamefont
  {Zyla}} \emph {et~al.} (\bibinfo {collaboration} {Particle Data Group}),\
  }\bibfield  {title} {\bibinfo {title} {{Review of Particle Physics}},\ }\href
  {https://doi.org/10.1093/ptep/ptaa104} {\bibfield  {journal} {\bibinfo
  {journal} {PTEP}\ }\textbf {\bibinfo {volume} {2020}},\ \bibinfo {pages}
  {083C01} (\bibinfo {year} {2020})}\BibitemShut {NoStop}%
\bibitem [{\citenamefont {Ethier}\ \emph {et~al.}(2021)\citenamefont {Ethier},
  \citenamefont {Magni}, \citenamefont {Maltoni}, \citenamefont {Mantani},
  \citenamefont {Nocera}, \citenamefont {Rojo}, \citenamefont {Slade},
  \citenamefont {Vryonidou},\ and\ \citenamefont {Zhang}}]{Ethier:2021bye}%
  \BibitemOpen
  \bibfield  {author} {\bibinfo {author} {\bibfnamefont {J.~J.}\ \bibnamefont
  {Ethier}}, \bibinfo {author} {\bibfnamefont {G.}~\bibnamefont {Magni}},
  \bibinfo {author} {\bibfnamefont {F.}~\bibnamefont {Maltoni}}, \bibinfo
  {author} {\bibfnamefont {L.}~\bibnamefont {Mantani}}, \bibinfo {author}
  {\bibfnamefont {E.~R.}\ \bibnamefont {Nocera}}, \bibinfo {author}
  {\bibfnamefont {J.}~\bibnamefont {Rojo}}, \bibinfo {author} {\bibfnamefont
  {E.}~\bibnamefont {Slade}}, \bibinfo {author} {\bibfnamefont
  {E.}~\bibnamefont {Vryonidou}},\ and\ \bibinfo {author} {\bibfnamefont
  {C.}~\bibnamefont {Zhang}} (\bibinfo {collaboration} {SMEFiT}),\ }\bibfield
  {title} {\bibinfo {title} {{Combined SMEFT interpretation of Higgs, diboson,
  and top quark data from the LHC}},\ }\href
  {https://doi.org/10.1007/JHEP11(2021)089} {\bibfield  {journal} {\bibinfo
  {journal} {JHEP}\ }\textbf {\bibinfo {volume} {11}},\ \bibinfo {pages}
  {089}},\ \Eprint {https://arxiv.org/abs/2105.00006} {arXiv:2105.00006
  [hep-ph]} \BibitemShut {NoStop}%
\bibitem [{\citenamefont {Brown}\ \emph {et~al.}(2021)\citenamefont {Brown},
  \citenamefont {Buckley}, \citenamefont {Englert}, \citenamefont {Ferrando},
  \citenamefont {Galler}, \citenamefont {Miller}, \citenamefont {Wanotayaroj},
  \citenamefont {Warrack},\ and\ \citenamefont {White}}]{Brown:2020sjx}%
  \BibitemOpen
  \bibfield  {author} {\bibinfo {author} {\bibfnamefont {S.}~\bibnamefont
  {Brown}}, \bibinfo {author} {\bibfnamefont {A.}~\bibnamefont {Buckley}},
  \bibinfo {author} {\bibfnamefont {C.}~\bibnamefont {Englert}}, \bibinfo
  {author} {\bibfnamefont {J.}~\bibnamefont {Ferrando}}, \bibinfo {author}
  {\bibfnamefont {P.}~\bibnamefont {Galler}}, \bibinfo {author} {\bibfnamefont
  {D.}~\bibnamefont {Miller}}, \bibinfo {author} {\bibfnamefont
  {C.}~\bibnamefont {Wanotayaroj}}, \bibinfo {author} {\bibfnamefont
  {N.}~\bibnamefont {Warrack}},\ and\ \bibinfo {author} {\bibfnamefont
  {C.}~\bibnamefont {White}},\ }\bibfield  {title} {\bibinfo {title} {{New
  results from TopFitter}},\ }\href {https://doi.org/10.22323/1.390.0322}
  {\bibfield  {journal} {\bibinfo  {journal} {PoS}\ }\textbf {\bibinfo {volume}
  {ICHEP2020}},\ \bibinfo {pages} {322} (\bibinfo {year} {2021})}\BibitemShut
  {NoStop}%
\bibitem [{\citenamefont {Ellis}\ \emph {et~al.}(2021)\citenamefont {Ellis},
  \citenamefont {Madigan}, \citenamefont {Mimasu}, \citenamefont {Sanz},\ and\
  \citenamefont {You}}]{Ellis:2020unq}%
  \BibitemOpen
  \bibfield  {author} {\bibinfo {author} {\bibfnamefont {J.}~\bibnamefont
  {Ellis}}, \bibinfo {author} {\bibfnamefont {M.}~\bibnamefont {Madigan}},
  \bibinfo {author} {\bibfnamefont {K.}~\bibnamefont {Mimasu}}, \bibinfo
  {author} {\bibfnamefont {V.}~\bibnamefont {Sanz}},\ and\ \bibinfo {author}
  {\bibfnamefont {T.}~\bibnamefont {You}},\ }\bibfield  {title} {\bibinfo
  {title} {{Top, Higgs, Diboson and Electroweak Fit to the Standard Model
  Effective Field Theory}},\ }\href {https://doi.org/10.1007/JHEP04(2021)279}
  {\bibfield  {journal} {\bibinfo  {journal} {JHEP}\ }\textbf {\bibinfo
  {volume} {04}},\ \bibinfo {pages} {279}},\ \Eprint
  {https://arxiv.org/abs/2012.02779} {arXiv:2012.02779 [hep-ph]} \BibitemShut
  {NoStop}%
\bibitem [{\citenamefont {Miralles}\ \emph {et~al.}(2021)\citenamefont
  {Miralles}, \citenamefont {L\'opez}, \citenamefont {Ll\'acer}, \citenamefont
  {Pe\~nuelas}, \citenamefont {Perell\'o},\ and\ \citenamefont
  {Vos}}]{Miralles:2021dyw}%
  \BibitemOpen
  \bibfield  {author} {\bibinfo {author} {\bibfnamefont {V.}~\bibnamefont
  {Miralles}}, \bibinfo {author} {\bibfnamefont {M.~M.}\ \bibnamefont
  {L\'opez}}, \bibinfo {author} {\bibfnamefont {M.~M.}\ \bibnamefont
  {Ll\'acer}}, \bibinfo {author} {\bibfnamefont {A.}~\bibnamefont
  {Pe\~nuelas}}, \bibinfo {author} {\bibfnamefont {M.}~\bibnamefont
  {Perell\'o}},\ and\ \bibinfo {author} {\bibfnamefont {M.}~\bibnamefont
  {Vos}},\ }\bibfield  {title} {\bibinfo {title} {{The top quark electro-weak
  couplings after LHC Run 2}},\ }\href@noop {} {\  (\bibinfo {year} {2021})},\
  \Eprint {https://arxiv.org/abs/2107.13917} {arXiv:2107.13917 [hep-ph]}
  \BibitemShut {NoStop}%
\bibitem [{\citenamefont {Bi\ss{}mann}\ \emph {et~al.}(2021)\citenamefont
  {Bi\ss{}mann}, \citenamefont {Grunwald}, \citenamefont {Hiller},\ and\
  \citenamefont {Kr\"oninger}}]{Bissmann:2020mfi}%
  \BibitemOpen
  \bibfield  {author} {\bibinfo {author} {\bibfnamefont {S.}~\bibnamefont
  {Bi\ss{}mann}}, \bibinfo {author} {\bibfnamefont {C.}~\bibnamefont
  {Grunwald}}, \bibinfo {author} {\bibfnamefont {G.}~\bibnamefont {Hiller}},\
  and\ \bibinfo {author} {\bibfnamefont {K.}~\bibnamefont {Kr\"oninger}},\
  }\bibfield  {title} {\bibinfo {title} {{Top and Beauty synergies in
  SMEFT-fits at present and future colliders}},\ }\href
  {https://doi.org/10.1007/JHEP06(2021)010} {\bibfield  {journal} {\bibinfo
  {journal} {JHEP}\ }\textbf {\bibinfo {volume} {06}},\ \bibinfo {pages}
  {010}},\ \Eprint {https://arxiv.org/abs/2012.10456} {arXiv:2012.10456
  [hep-ph]} \BibitemShut {NoStop}%
\bibitem [{\citenamefont {Korchin}\ and\ \citenamefont
  {Kovalchuk}(2013)}]{Korchin:2013ifa}%
  \BibitemOpen
  \bibfield  {author} {\bibinfo {author} {\bibfnamefont {A.~Y.}\ \bibnamefont
  {Korchin}}\ and\ \bibinfo {author} {\bibfnamefont {V.~A.}\ \bibnamefont
  {Kovalchuk}},\ }\bibfield  {title} {\bibinfo {title} {{Polarization effects
  in the Higgs boson decay to $\gamma Z$ and test of $CP$ and $CPT$
  symmetries}},\ }\href {https://doi.org/10.1103/PhysRevD.88.036009} {\bibfield
   {journal} {\bibinfo  {journal} {Phys. Rev. D}\ }\textbf {\bibinfo {volume}
  {88}},\ \bibinfo {pages} {036009} (\bibinfo {year} {2013})},\ \Eprint
  {https://arxiv.org/abs/1303.0365} {arXiv:1303.0365 [hep-ph]} \BibitemShut
  {NoStop}%
\bibitem [{\citenamefont {Korchin}\ and\ \citenamefont
  {Kovalchuk}(2014)}]{Korchin:2014kha}%
  \BibitemOpen
  \bibfield  {author} {\bibinfo {author} {\bibfnamefont {A.~Y.}\ \bibnamefont
  {Korchin}}\ and\ \bibinfo {author} {\bibfnamefont {V.~A.}\ \bibnamefont
  {Kovalchuk}},\ }\bibfield  {title} {\bibinfo {title} {{Angular distribution
  and forward\textendash{}backward asymmetry of the Higgs-boson decay to photon
  and lepton pair}},\ }\href {https://doi.org/10.1140/epjc/s10052-014-3141-7}
  {\bibfield  {journal} {\bibinfo  {journal} {Eur. Phys. J. C}\ }\textbf
  {\bibinfo {volume} {74}},\ \bibinfo {pages} {3141} (\bibinfo {year}
  {2014})},\ \Eprint {https://arxiv.org/abs/1408.0342} {arXiv:1408.0342
  [hep-ph]} \BibitemShut {NoStop}%
\bibitem [{\citenamefont {Korchin}\ and\ \citenamefont
  {Kovalchuk}(2016)}]{Korchin:2016rsf}%
  \BibitemOpen
  \bibfield  {author} {\bibinfo {author} {\bibfnamefont {A.~Y.}\ \bibnamefont
  {Korchin}}\ and\ \bibinfo {author} {\bibfnamefont {V.~A.}\ \bibnamefont
  {Kovalchuk}},\ }\bibfield  {title} {\bibinfo {title} {{Decay of the Higgs
  boson to $\tau^- \tau^+$ and non-Hermiticity of the Yukawa interaction}},\
  }\href {https://doi.org/10.1103/PhysRevD.94.076003} {\bibfield  {journal}
  {\bibinfo  {journal} {Phys. Rev. D}\ }\textbf {\bibinfo {volume} {94}},\
  \bibinfo {pages} {076003} (\bibinfo {year} {2016})},\ \Eprint
  {https://arxiv.org/abs/1607.02827} {arXiv:1607.02827 [hep-ph]} \BibitemShut
  {NoStop}%
\bibitem [{\citenamefont {Chen}\ and\ \citenamefont {Wu}(2019)}]{Chen:2017nxp}%
  \BibitemOpen
  \bibfield  {author} {\bibinfo {author} {\bibfnamefont {X.}~\bibnamefont
  {Chen}}\ and\ \bibinfo {author} {\bibfnamefont {Y.}~\bibnamefont {Wu}},\
  }\bibfield  {title} {\bibinfo {title} {{Probing the CP-Violation effects in
  the $h\tau\tau$ coupling at the LHC}},\ }\href
  {https://doi.org/10.1016/j.physletb.2019.01.038} {\bibfield  {journal}
  {\bibinfo  {journal} {Phys. Lett. B}\ }\textbf {\bibinfo {volume} {790}},\
  \bibinfo {pages} {332} (\bibinfo {year} {2019})},\ \Eprint
  {https://arxiv.org/abs/1708.02882} {arXiv:1708.02882 [hep-ph]} \BibitemShut
  {NoStop}%
\bibitem [{\citenamefont {Sirunyan}\ \emph {et~al.}(2020)\citenamefont
  {Sirunyan} \emph {et~al.}}]{CMS:2020cga}%
  \BibitemOpen
  \bibfield  {author} {\bibinfo {author} {\bibfnamefont {A.~M.}\ \bibnamefont
  {Sirunyan}} \emph {et~al.} (\bibinfo {collaboration} {CMS}),\ }\bibfield
  {title} {\bibinfo {title} {{Measurements of $\mathrm{t\bar{t}}H$ Production
  and the CP Structure of the Yukawa Interaction between the Higgs Boson and
  Top Quark in the Diphoton Decay Channel}},\ }\href
  {https://doi.org/10.1103/PhysRevLett.125.061801} {\bibfield  {journal}
  {\bibinfo  {journal} {Phys. Rev. Lett.}\ }\textbf {\bibinfo {volume} {125}},\
  \bibinfo {pages} {061801} (\bibinfo {year} {2020})},\ \Eprint
  {https://arxiv.org/abs/2003.10866} {arXiv:2003.10866 [hep-ex]} \BibitemShut
  {NoStop}%
\bibitem [{\citenamefont {Bernreuther}\ \emph
  {et~al.}(2005{\natexlab{a}})\citenamefont {Bernreuther}, \citenamefont
  {Bonciani}, \citenamefont {Gehrmann}, \citenamefont {Heinesch}, \citenamefont
  {Leineweber}, \citenamefont {Mastrolia},\ and\ \citenamefont
  {Remiddi}}]{Bernreuther:2004ih}%
  \BibitemOpen
  \bibfield  {author} {\bibinfo {author} {\bibfnamefont {W.}~\bibnamefont
  {Bernreuther}}, \bibinfo {author} {\bibfnamefont {R.}~\bibnamefont
  {Bonciani}}, \bibinfo {author} {\bibfnamefont {T.}~\bibnamefont {Gehrmann}},
  \bibinfo {author} {\bibfnamefont {R.}~\bibnamefont {Heinesch}}, \bibinfo
  {author} {\bibfnamefont {T.}~\bibnamefont {Leineweber}}, \bibinfo {author}
  {\bibfnamefont {P.}~\bibnamefont {Mastrolia}},\ and\ \bibinfo {author}
  {\bibfnamefont {E.}~\bibnamefont {Remiddi}},\ }\bibfield  {title} {\bibinfo
  {title} {{Two-loop QCD corrections to the heavy quark form-factors: The
  Vector contributions}},\ }\href
  {https://doi.org/10.1016/j.nuclphysb.2004.10.059} {\bibfield  {journal}
  {\bibinfo  {journal} {Nucl. Phys. B}\ }\textbf {\bibinfo {volume} {706}},\
  \bibinfo {pages} {245} (\bibinfo {year} {2005}{\natexlab{a}})},\ \Eprint
  {https://arxiv.org/abs/hep-ph/0406046} {arXiv:hep-ph/0406046} \BibitemShut
  {NoStop}%
\bibitem [{\citenamefont {Bernreuther}\ \emph
  {et~al.}(2005{\natexlab{b}})\citenamefont {Bernreuther}, \citenamefont
  {Bonciani}, \citenamefont {Gehrmann}, \citenamefont {Heinesch}, \citenamefont
  {Leineweber}, \citenamefont {Mastrolia},\ and\ \citenamefont
  {Remiddi}}]{Bernreuther:2004th}%
  \BibitemOpen
  \bibfield  {author} {\bibinfo {author} {\bibfnamefont {W.}~\bibnamefont
  {Bernreuther}}, \bibinfo {author} {\bibfnamefont {R.}~\bibnamefont
  {Bonciani}}, \bibinfo {author} {\bibfnamefont {T.}~\bibnamefont {Gehrmann}},
  \bibinfo {author} {\bibfnamefont {R.}~\bibnamefont {Heinesch}}, \bibinfo
  {author} {\bibfnamefont {T.}~\bibnamefont {Leineweber}}, \bibinfo {author}
  {\bibfnamefont {P.}~\bibnamefont {Mastrolia}},\ and\ \bibinfo {author}
  {\bibfnamefont {E.}~\bibnamefont {Remiddi}},\ }\bibfield  {title} {\bibinfo
  {title} {{Two-loop QCD corrections to the heavy quark form-factors: Axial
  vector contributions}},\ }\href
  {https://doi.org/10.1016/j.nuclphysb.2005.01.035} {\bibfield  {journal}
  {\bibinfo  {journal} {Nucl. Phys. B}\ }\textbf {\bibinfo {volume} {712}},\
  \bibinfo {pages} {229} (\bibinfo {year} {2005}{\natexlab{b}})},\ \Eprint
  {https://arxiv.org/abs/hep-ph/0412259} {arXiv:hep-ph/0412259} \BibitemShut
  {NoStop}%
\bibitem [{\citenamefont {Bernreuther}\ \emph
  {et~al.}(2005{\natexlab{c}})\citenamefont {Bernreuther}, \citenamefont
  {Bonciani}, \citenamefont {Gehrmann}, \citenamefont {Heinesch}, \citenamefont
  {Leineweber},\ and\ \citenamefont {Remiddi}}]{Bernreuther:2005rw}%
  \BibitemOpen
  \bibfield  {author} {\bibinfo {author} {\bibfnamefont {W.}~\bibnamefont
  {Bernreuther}}, \bibinfo {author} {\bibfnamefont {R.}~\bibnamefont
  {Bonciani}}, \bibinfo {author} {\bibfnamefont {T.}~\bibnamefont {Gehrmann}},
  \bibinfo {author} {\bibfnamefont {R.}~\bibnamefont {Heinesch}}, \bibinfo
  {author} {\bibfnamefont {T.}~\bibnamefont {Leineweber}},\ and\ \bibinfo
  {author} {\bibfnamefont {E.}~\bibnamefont {Remiddi}},\ }\bibfield  {title}
  {\bibinfo {title} {{Two-loop QCD corrections to the heavy quark form-factors:
  Anomaly contributions}},\ }\href
  {https://doi.org/10.1016/j.nuclphysb.2005.06.025} {\bibfield  {journal}
  {\bibinfo  {journal} {Nucl. Phys. B}\ }\textbf {\bibinfo {volume} {723}},\
  \bibinfo {pages} {91} (\bibinfo {year} {2005}{\natexlab{c}})},\ \Eprint
  {https://arxiv.org/abs/hep-ph/0504190} {arXiv:hep-ph/0504190} \BibitemShut
  {NoStop}%
\bibitem [{\citenamefont {Czarnecki}\ \emph {et~al.}(1996)\citenamefont
  {Czarnecki}, \citenamefont {Krause},\ and\ \citenamefont
  {Marciano}}]{Czarnecki:1995sz}%
  \BibitemOpen
  \bibfield  {author} {\bibinfo {author} {\bibfnamefont {A.}~\bibnamefont
  {Czarnecki}}, \bibinfo {author} {\bibfnamefont {B.}~\bibnamefont {Krause}},\
  and\ \bibinfo {author} {\bibfnamefont {W.~J.}\ \bibnamefont {Marciano}},\
  }\bibfield  {title} {\bibinfo {title} {{Electroweak corrections to the muon
  anomalous magnetic moment}},\ }\href
  {https://doi.org/10.1103/PhysRevLett.76.3267} {\bibfield  {journal} {\bibinfo
   {journal} {Phys. Rev. Lett.}\ }\textbf {\bibinfo {volume} {76}},\ \bibinfo
  {pages} {3267} (\bibinfo {year} {1996})},\ \Eprint
  {https://arxiv.org/abs/hep-ph/9512369} {arXiv:hep-ph/9512369} \BibitemShut
  {NoStop}%
\bibitem [{\citenamefont {Hollik}(1990)}]{Hollik:1988ii}%
  \BibitemOpen
  \bibfield  {author} {\bibinfo {author} {\bibfnamefont {W.~F.~L.}\
  \bibnamefont {Hollik}},\ }\bibfield  {title} {\bibinfo {title} {{Radiative
  Corrections in the Standard Model and their Role for Precision Tests of the
  Electroweak Theory}},\ }\href {https://doi.org/10.1002/prop.2190380302}
  {\bibfield  {journal} {\bibinfo  {journal} {Fortsch. Phys.}\ }\textbf
  {\bibinfo {volume} {38}},\ \bibinfo {pages} {165} (\bibinfo {year}
  {1990})}\BibitemShut {NoStop}%
\bibitem [{\citenamefont {Bernabeu}\ \emph {et~al.}(1997)\citenamefont
  {Bernabeu}, \citenamefont {Vidal},\ and\ \citenamefont
  {Gonzalez-Sprinberg}}]{Bernabeu:1997je}%
  \BibitemOpen
  \bibfield  {author} {\bibinfo {author} {\bibfnamefont {J.}~\bibnamefont
  {Bernabeu}}, \bibinfo {author} {\bibfnamefont {J.}~\bibnamefont {Vidal}},\
  and\ \bibinfo {author} {\bibfnamefont {G.~A.}\ \bibnamefont
  {Gonzalez-Sprinberg}},\ }\bibfield  {title} {\bibinfo {title} {{The Weak
  magnetic moment of heavy quarks}},\ }\href
  {https://doi.org/10.1016/S0370-2693(97)00185-8} {\bibfield  {journal}
  {\bibinfo  {journal} {Phys. Lett. B}\ }\textbf {\bibinfo {volume} {397}},\
  \bibinfo {pages} {255} (\bibinfo {year} {1997})},\ \Eprint
  {https://arxiv.org/abs/hep-ph/9702222} {arXiv:hep-ph/9702222} \BibitemShut
  {NoStop}%
\bibitem [{\citenamefont {Bernreuther}\ \emph
  {et~al.}(2005{\natexlab{d}})\citenamefont {Bernreuther}, \citenamefont
  {Bonciani}, \citenamefont {Gehrmann}, \citenamefont {Heinesch}, \citenamefont
  {Leineweber}, \citenamefont {Mastrolia},\ and\ \citenamefont
  {Remiddi}}]{Bernreuther:2005gq}%
  \BibitemOpen
  \bibfield  {author} {\bibinfo {author} {\bibfnamefont {W.}~\bibnamefont
  {Bernreuther}}, \bibinfo {author} {\bibfnamefont {R.}~\bibnamefont
  {Bonciani}}, \bibinfo {author} {\bibfnamefont {T.}~\bibnamefont {Gehrmann}},
  \bibinfo {author} {\bibfnamefont {R.}~\bibnamefont {Heinesch}}, \bibinfo
  {author} {\bibfnamefont {T.}~\bibnamefont {Leineweber}}, \bibinfo {author}
  {\bibfnamefont {P.}~\bibnamefont {Mastrolia}},\ and\ \bibinfo {author}
  {\bibfnamefont {E.}~\bibnamefont {Remiddi}},\ }\bibfield  {title} {\bibinfo
  {title} {{QCD corrections to static heavy quark form-factors}},\ }\href
  {https://doi.org/10.1103/PhysRevLett.95.261802} {\bibfield  {journal}
  {\bibinfo  {journal} {Phys. Rev. Lett.}\ }\textbf {\bibinfo {volume} {95}},\
  \bibinfo {pages} {261802} (\bibinfo {year} {2005}{\natexlab{d}})},\ \Eprint
  {https://arxiv.org/abs/hep-ph/0509341} {arXiv:hep-ph/0509341} \BibitemShut
  {NoStop}%
\end{thebibliography}%

\end{document}